\documentclass[reprint,amsmath,amssymb,aps,twocolumn]{revtex4-2}

\usepackage{graphicx}
\usepackage{mathtools}
\usepackage{dcolumn}
\usepackage{bm}

\usepackage[]{hyperref}
\usepackage{color} % color
\usepackage{amsmath}

\begin{document}

\title{Observation of Highly Correlated Ultrabright Biphotons Through Increased Atomic Ensemble Density in Spontaneous Four-Wave Mixing}

\author{Jiun-Shiuan Shiu,$^{1,2}$ Zi-Yu Liu,$^{1,2}$ Chin-Yao Cheng,$^{1,2}$ Yu-Chiao Huang,$^{1,2}$ Ite A. Yu,$^{3,4}$ Ying-Cheng Chen,$^{5}$ Chih-Sung Chuu,$^{3,4}$ Che-Ming Li,$^{2,6}$ Shiang-Yu Wang,$^7$ and Yong-Fan Chen$^{1,2,}$}

\email{yfchen@mail.ncku.edu.tw}

\affiliation{
$^1$Department of Physics, National Cheng Kung University, Tainan 70101, Taiwan\\ 
$^2$Center for Quantum Frontiers of Research $\&$ Technology, Tainan 70101, Taiwan\\ 
$^3$Center for Quantum Science and Technology, National Tsing Hua University, Hsinchu 30013, Taiwan\\
$^4$Department of Physics, National Tsing Hua University, Hsinchu 30013, Taiwan\\
$^5$Institute of Atomic and Molecular Sciences, Academia Sinica, Taipei 10617, Taiwan\\
$^6$Department of Engineering Science, National Cheng Kung University, Tainan 70101, Taiwan\\
$^7$Institute of Astronomy and Astrophysics, Academia Sinica, Taipei 10617, Taiwan
}

\date{February 10, 2024}
%\date{\today}

%%%%%%%%%%%%%%%%%%%%%%%%%%%%%%%%%%%%%%%%%%%%%%%%%%%%%%%%%%%%%%%%%%%%%%%%%%%%%%%%%%%%%%%%%%%%%%%%%%%%%
%%%%%%%%%%%%%%%%%%%%%%%%%%%%%%%%%%%%%%%%%%%%%%%%%%%%%%%%%%%%%%%%%%%%%%%%%%%%%%%%%%%%%%%%%%%%%%%%%%%%%

\begin{abstract}

The pairing ratio, a crucial metric assessing a biphoton source's ability to generate correlated photon pairs, remains underexplored despite theoretical predictions. This study presents experimental findings on the pairing ratio, utilizing a double-$\Lambda$ spontaneous four-wave mixing biphoton source in cold atoms. At an optical depth (OD) of 20, we achieved an ultrahigh biphoton generation rate of up to $1.3\times10^7$ per second, with a successful pairing ratio of 61\%. Increasing the OD to 120 significantly improved the pairing ratio to 89\%, while maintaining a consistent biphoton generation rate. This achievement, marked by high generation rates and robust biphoton pairing, holds great promise for advancing efficiency in quantum communication and information processing. Additionally, in a scenario with a lower biphoton generation rate of $5.0 \times 10^4$ per second, we attained an impressive signal-to-background ratio of 241 for the biphoton wavepacket, surpassing the Cauchy-Schwarz criterion by approximately $1.5\times10^4$ times.

\end{abstract}

%%%%%%%%%%%%%%%%%%%%%%%%%%%%%%%%%%%%%%%%%%%%%%%%%%%%%%%%%%%%%%%%%%%%%%%%%%%%%%%%%%%%%%%%%%%%%%%%%%%%%
%%%%%%%%%%%%%%%%%%%%%%%%%%%%%%%%%%%%%%%%%%%%%%%%%%%%%%%%%%%%%%%%%%%%%%%%%%%%%%%%%%%%%%%%%%%%%%%%%%%%%

\maketitle

%%%%%%%%%%%%%%%%%%%%%%%%%%%%%%%%%%%%%%%%%%%%%%%%%%%%%%%%%%%%%%%%%%%%%%%%%%%%%%%%%%%%%%%%%%%%%%%%%%%%%
%%%%%%%%%%%%%%%%%%%%%%%%%%%%%%%%%%%%%%%%%%%%%%%%%%%%%%%%%%%%%%%%%%%%%%%%%%%%%%%%%%%%%%%%%%%%%%%%%%%%%

\newcommand{\FigOne}{
    \begin{figure}[t]
    \centering
    \includegraphics[width = 8.8 cm]{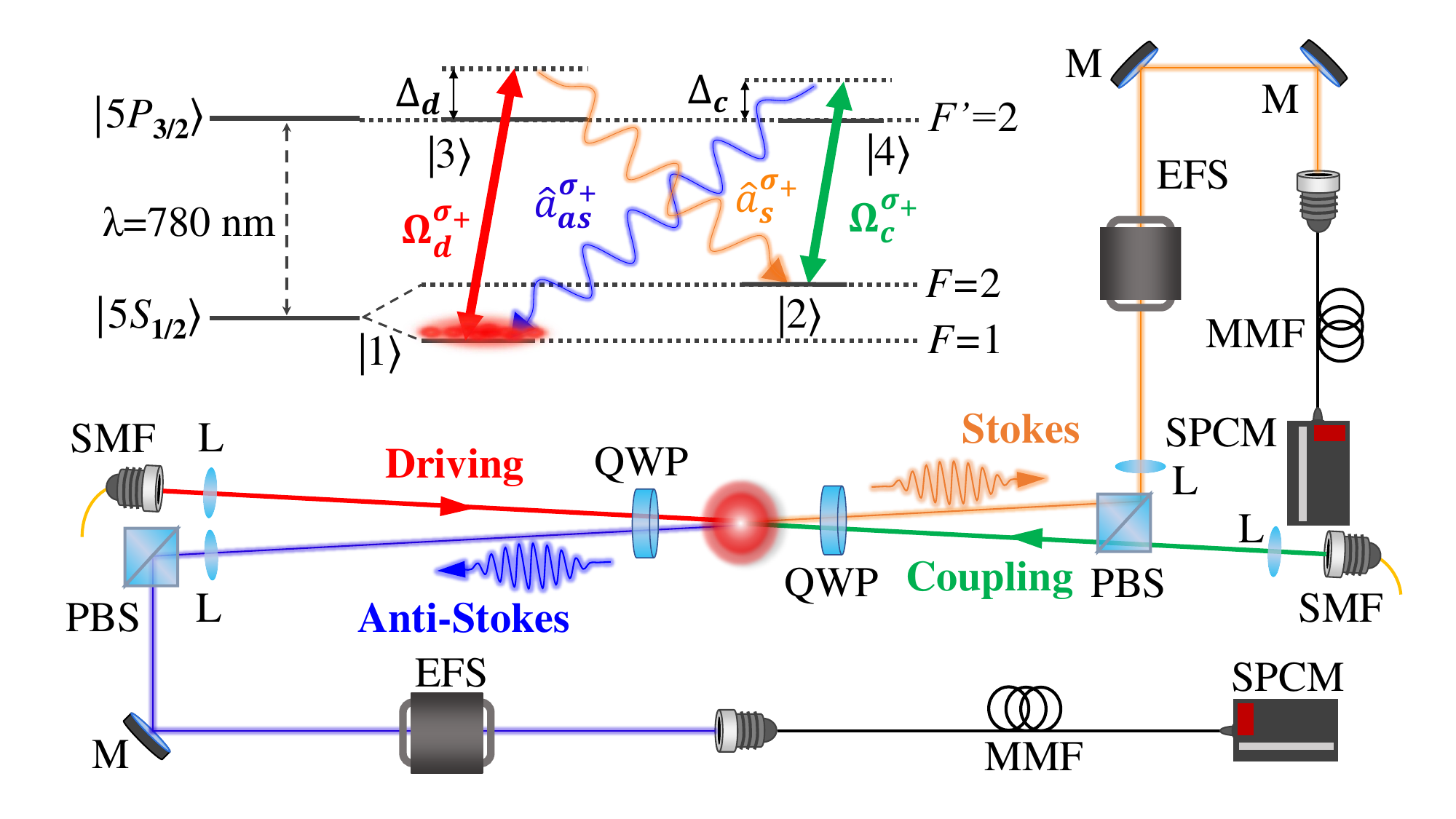}
    \caption{
Diagram of the double-$\Lambda$ SFWM system and experimental setup. M, mirror; L, lens; PBS, polarizing beam splitter; QWP, quarter-wave plate; SMF, single-mode fiber; MMF, multimode fiber; EFS, etalon filter set; SPCM, single-photon counting module. The inset shows the relevant energy levels of the $^{87}$Rb atom.
}
    \label{fig:Experimental setup}
    \end{figure}
}

\newcommand{\FigTwo}{
    \begin{figure}[t]
    \centering
    \includegraphics[width = 8.8 cm]{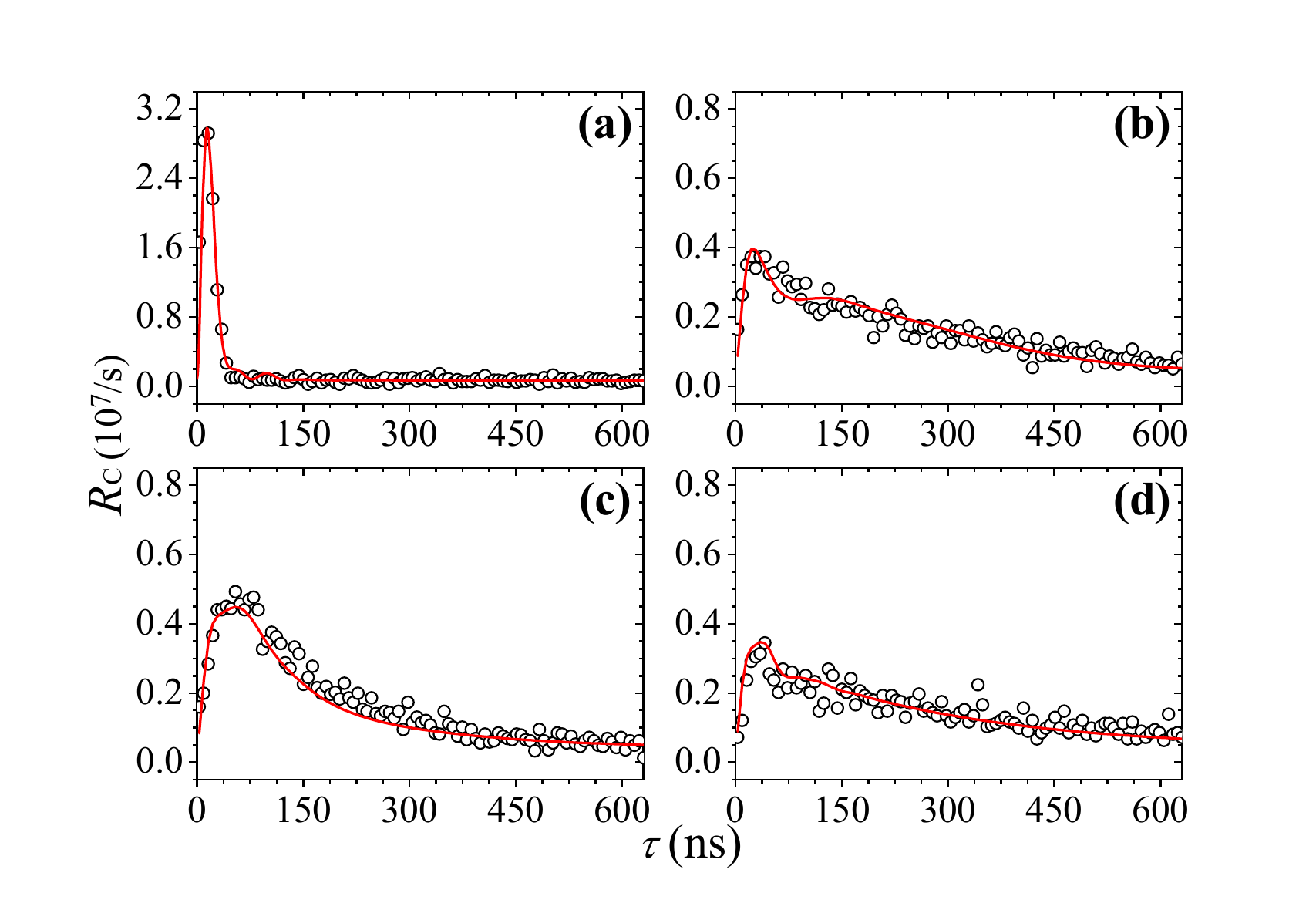}
    \caption{
Biphotons with controllable bandwidth. The red lines represent the theoretical curves, while the black circles indicate the experimental data points. The time bin for detecting the anti-Stokes photons is $\Delta\tau=6.4$ ns. Other parameters are ${\rm OD}=15$, $\Omega_d=1\Gamma$, $\Delta_d=10\Gamma$, $\gamma_{21}=0.001\Gamma$, $\Delta kL=0.37\pi$,
(a) $\Omega_c=4\Gamma$, $\Delta_c=0\Gamma$, 
(b) $\Omega_c=1\Gamma$, $\Delta_c=0\Gamma$, 
(c) $\Omega_c=1\Gamma$, $\Delta_c=1\Gamma$, 
(d) $\Omega_c=1\Gamma$, $\Delta_c=3\Gamma$.
}
    \label{fig:Slow light regime}
    \end{figure}
}

\newcommand{\FigThree}{
    \begin{figure}[t]
    \centering
    \includegraphics[width = 8.6 cm]{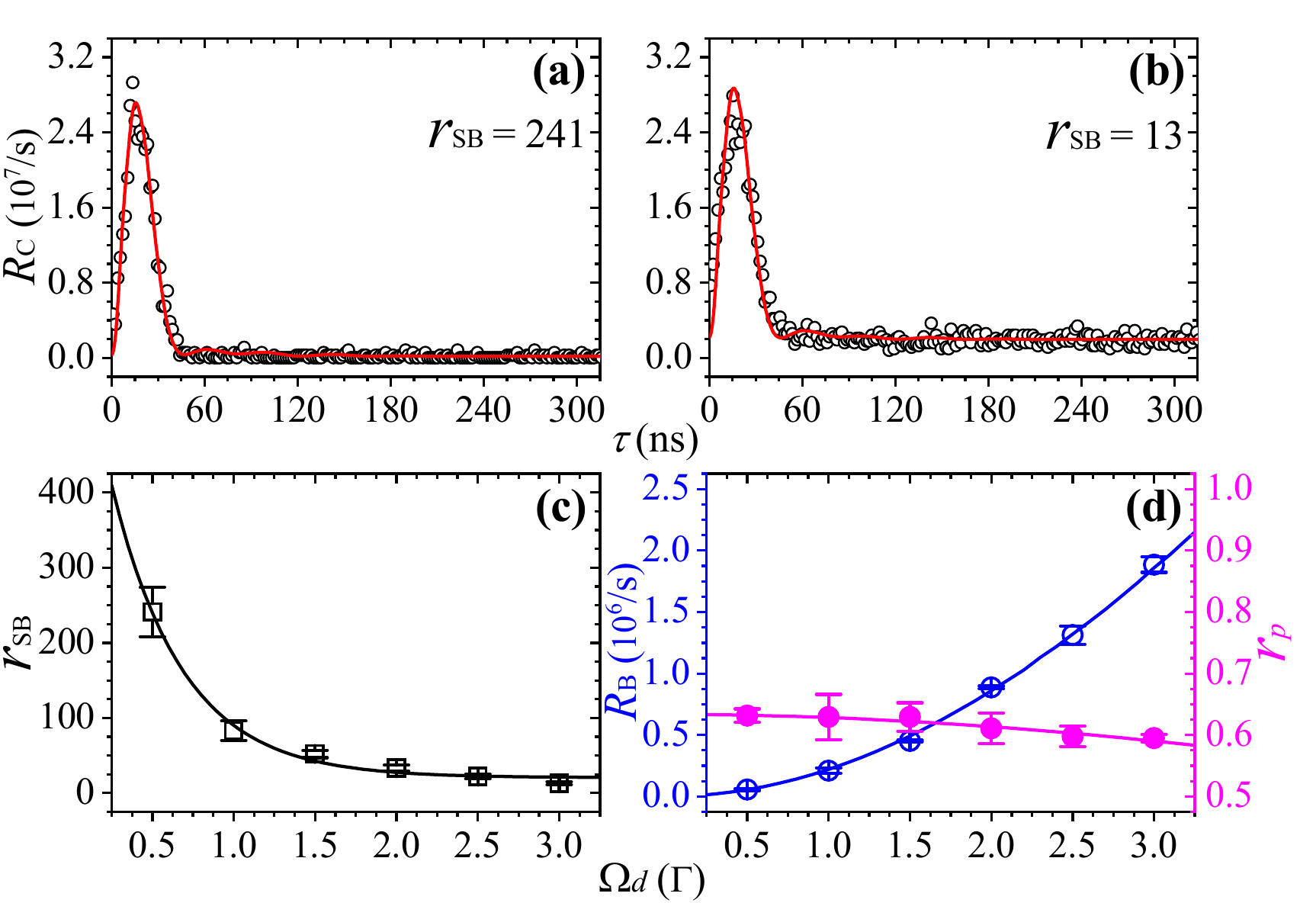}
    \caption{
High-purity biphotons. The red lines represent the theoretical curves, while the black circles indicate the experimental data points. The time bin for detecting the anti-Stokes photons is $\Delta\tau=1.6$ ns. The remaining parameters are set to ${\rm OD}=10$, $\Omega_c=4\Gamma$, $\Delta_d=10\Gamma$, $\gamma_{21}=0.001\Gamma$, $\Delta kL=0.37\pi$, with (a) $\Omega_d=0.5\Gamma$ and (b) $\Omega_d=3\Gamma$. (c) The peak signal-to-background ratio $r_{\rm SB}$ versus $\Omega_d$. The black squares represent the experimental data, and the black line is the curve fitted to these experimental data points. (d) The biphoton generation rate $R_{\rm B}$ and pairing ratio $r_p$ as a function of $\Omega_d$. The experimental data points for $R_{\rm B}$ and $r_p$ are represented by the unfilled blue and solid magenta circles, respectively. The theoretical curves for $R_{\rm B}$ and $r_p$ are depicted by the blue and magenta lines, respectively.
}
    \label{fig:high purity biphoton} 
    \end{figure}
}

\newcommand{\FigFour}{
    \begin{figure}[t]
    \centering
    \includegraphics[width = 8.6 cm]{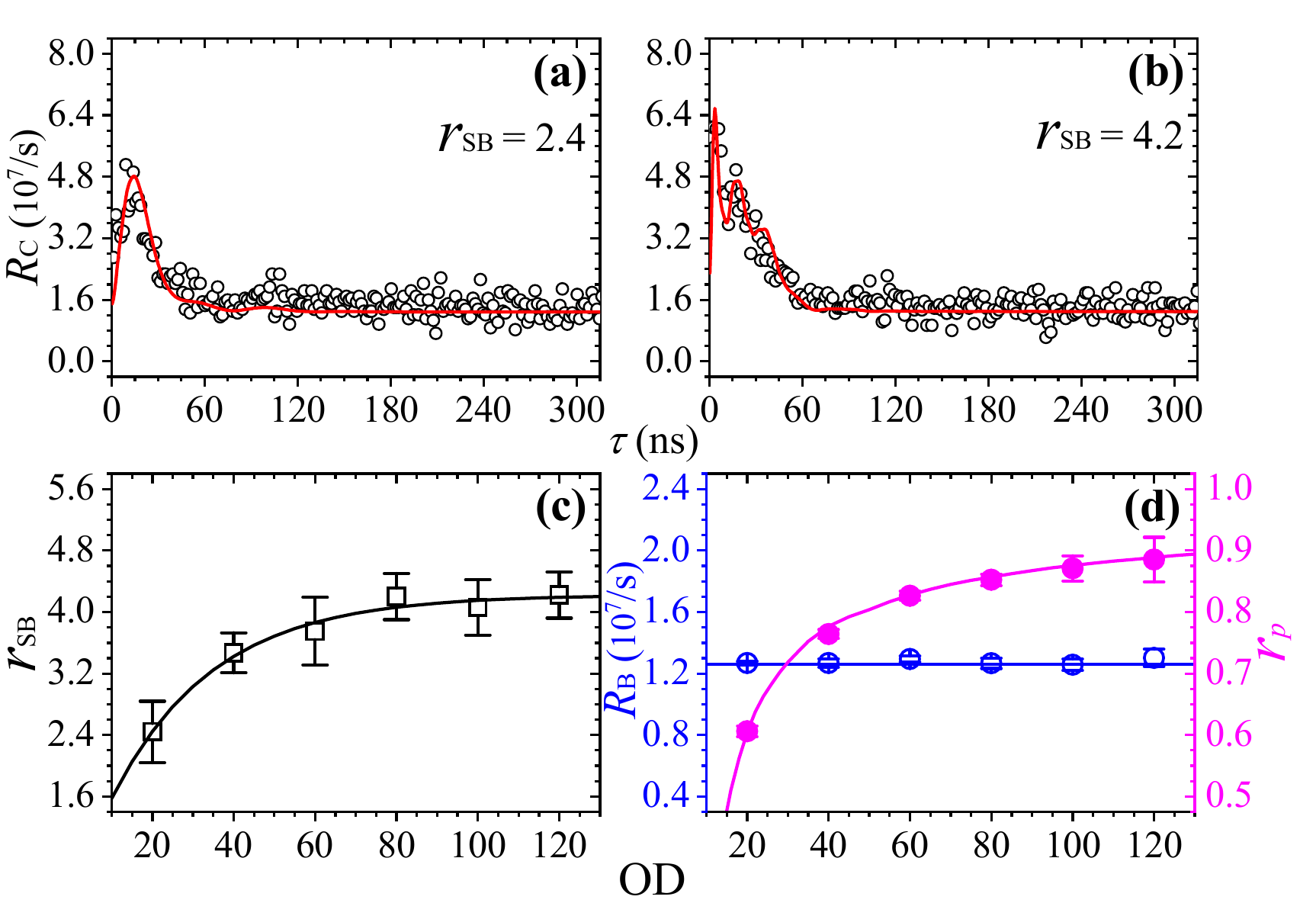}
    \caption{
Highly correlated ultrabright biphotons. The red lines represent the theoretical curves, while the black circles indicate the experimental data points. The time bin for detecting the anti-Stokes photons is $\Delta\tau=1.6$ ns. The remaining parameters are set to $\Omega_d=3\Gamma$, $\gamma_{21}=0.001\Gamma$, $\Delta kL=0.37\pi$, with (a) ${\rm OD}=20$, $\Omega_c=4\Gamma$, $\Delta_d=5\Gamma$, and (b) ${\rm OD}=120$, $\Omega_c=8.8\Gamma$, $\Delta_d=14.9\Gamma$. (c) The peak signal-to-background ratio $r_{\rm SB}$ versus OD. The black squares represent the experimental data, and the black line is the curve fitted to these experimental data points. (d) The biphoton generation rate $R_{\rm B}$ and pairing ratio $r_p$ as a function of OD. The experimental data points for $R_{\rm B}$ and $r_p$ are represented by the unfilled blue and solid magenta circles, respectively. The theoretical curves for $R_{\rm B}$ and $r_p$ are depicted by the blue and magenta lines, respectively.
}
    \label{fig:pairing ratio} 
    \end{figure}
}
%%%%%%%%%%%%%%%%%%%%%%%%%%%%%%%%%%%%%%%%%%%%%%%%%%%%%%%%%%%%%%%%%%%%%%%%%%%%%%%%%%%%%%%%%%%%%%%%%%%%%
%%%%%%%%%%%%%%%%%%%%%%%%%%%%%%%%%%%%%%%%%%%%%%%%%%%%%%%%%%%%%%%%%%%%%%%%%%%%%%%%%%%%%%%%%%%%%%%%%%%%%

%\section{Introduction} %\label{sec:Introduction}

\textit{Introduction}.--Temporally correlated biphotons have recently garnered considerable attention in the fields of optical quantum computing and quantum communication, thanks to their exceptional nonclassical properties. Of particular significance is their role as heralded single-photon sources, which have found applications in diverse domains, including quantum cryptography~\cite{cryptography1, cryptography2, cryptography3, cryptography4}, quantum metrology~\cite{metrology1, metrology2, metrology3, metrology4}, and quantum imaging~\cite{imaging1, imaging2, imaging3, imaging4}. Among the various biphoton sources, the spontaneous four-wave mixing (SFWM) mechanism, distinguished by its operation near atomic resonance, stands out for its ability to conveniently manipulate bandwidth and serves as a bridge between different quantum devices, thus attracting significant interest.

The operation of SFWM near resonance accommodates diverse energy level configurations, including the double-$\Lambda$ scheme~\cite{Lambda1, Lambda2, Lambda3} and cascade-transition scheme~\cite{cascade1, cascade2, cascade3}. This proximity to atomic resonance enables the generation of bright biphotons with low optical power~\cite{RB1, RB2}, as well as the production of narrowband biphotons~\cite{bandwidth2, bandwidth5}. Especially for the double-$\Lambda$ scheme, characterized by its intrinsic $\Lambda$-type electromagnetically induced transparency (EIT) structure~\cite{EIT1, EIT2, EIT3, EIT4, EIT5}, it not only significantly suppresses the generation of noise photons~\cite{QM1, QM2, QFC, QM3} but also provides a wide bandwidth tuning capability~\cite{bandwidth1, bandwidth3, bandwidth4}. This facilitates its direct application in conjunction with quantum devices~\cite{device1, device2, FWM, HOM} or reshaping of the biphoton waveforms~\cite{modulation1, modulation2, modulation3}. Furthermore, the $\Lambda$ structure supports convenient implementation in two- or three-level atomic systems~\cite{twolevel1, twolevel2, twolevel3, twolevel4, twolevel5}. 

Despite the numerous remarkable achievements of the double-$\Lambda$ SFWM scheme, there is a frequently overlooked concern: the limited atomic density hinders spontaneously emitted photons from achieving perfect coherence through the four-wave mixing (FWM) process~\cite{FWM1, FWM2}. Incoherent emissions not involved in FWM processes can significantly diminish the biphoton pairing capacity, referred to as the pairing ratio. Unfortunately, while the Heisenberg--Langevin operator theory~\cite{Kolchin} is predictive in understanding the concept of the pairing ratio, there is currently a lack of relevant research and investigation in experimental studies.

In this Letter, we present the thorough investigation of the pairing ratio using the double-$\Lambda$ SFWM in a cold $^{87}$Rb ensemble. This configuration, chosen for its inherent EIT effect, allows for easy control of the biphoton bandwidth. We achieved an exceptionally high biphoton generation rate of approximately $1.3\times10^7$ per second at an optical depth (OD) of 20, with the pairing ratio measured at 0.61. We also illustrated that this ratio can be enhanced by increasing OD and achieved the highest pairing ratio of 0.89 at an OD of 120. Additionally, at a relatively low biphoton generation rate of $5.0 \times 10^4$ per second, the signal-to-background ratio for the biphoton wavepacket reached 241, exceeding the Cauchy-Schwarz criterion by approximately $1.5\times10^4$ times.

%%%%%%%%
\FigOne
%%%%%%%%

%%%%%%%%%%%%%%%%%%%%%%%%%%%%%%%%%%%%%%%%%%%%%%%%%%%%%%%%%%%%%%%%%%%%%%%%%%%%%%%%%%%%%%%%%%%%%%%%%%%%%
%%%%%%%%%%%%%%%%%%%%%%%%%%%%%%%%%%%%%%%%%%%%%%%%%%%%%%%%%%%%%%%%%%%%%%%%%%%%%%%%%%%%%%%%%%%%%%%%%%%%%

%\section{Experimental Details} %\label{sec:Experiment}

\textit{Experimental setup}.--We trapped cold $^{87}$Rb atoms using a standard magneto-optical trap. After optically pumping them to the ground state $|5S_{1/2}, F=1\rangle$, as illustrated in Fig. \ref{fig:Experimental setup}, we irradiated the atomic ensemble with a far-detuned driving field characterized by a Rabi frequency $\Omega_d$ and a nearly resonant coupling field denoted by $\Omega_c$. Synchronization between these fields was achieved through injection locking with an external cavity diode laser (not shown). The driving field, with detuning $\Delta_d$, operated on the $\sigma^+$-transition $|5S_{1/2}, F=1\rangle \rightarrow |5P_{3/2}, F=2\rangle$, effectively suppressing incoherent fluorescence from one-photon absorption. This allowed for primary emission of Stokes photons via spontaneous Raman scattering~\cite{SRS}. Subsequently, the nearly resonant coupling field, with detuning $\Delta_c$, acted on the $\sigma^+$-transition $|5S_{1/2}, F=2\rangle \rightarrow |5P_{3/2}, F=2\rangle$, inducing the emission of anti-Stokes photons. We used an elongated atomic ensemble to enhance the FWM effect and boosting specific direction scattering probability.

In the SFWM experiment, biphotons were generated using a 10-$\mu$s driving pulse in each 2.5-ms cycle. To prevent laser leakage, we adopted a backward configuration where the driving and coupling beams counter-propagated, each with $1/e^2$ full widths of 250 and 310 $\mu$m. The corresponding optical powers of the 1$\Gamma$ driving and coupling fields were approximately 7.5 and 11.1 $\mu$W. Both fields drove $\sigma^+$-transitions, resulting in generated photon pairs exhibiting $\sigma^+$ polarization, propagating in opposite directions and passing through respective etalon filter sets (EFS). The intersection angle between the Stokes (anti-Stokes) and driving (coupling) beams was set at 1.7$^\circ$ for our experiment. Each EFS consisted of two etalons, each with an extinction ratio of roughly 30 dB and an approximate bandwidth of 100 MHz, separated by an optical isolator. The total extinction ratios of the Stokes and anti-Stokes channels were 114 dB and 124 dB, respectively. Biphotons were detected using fiber-coupled single-photon counting modules (SPCM-AQRH-13-FC). Upon detection, an 8-ns pulse was emitted from the SPCM toward the time-of-flight multiscaler (TOF, MCS6A-4T8, not shown). In the coincidence count experiment, we measured the time difference between Stokes and anti-Stokes photons. When generated as a correlated pair, they arrived at the SPCMs within the correlation time, contributing to a nonflat biphoton wavepacket. The TOF generated a histogram of coincident counts based on these data points, providing insight into the source of biphotons.

%%%%%%%%%%%%%%%%%%%%%%%%%%%%%%%%%%%%%%%%%%%%%%%%%%%%%%%%%%%%%%%%%%%%%%%%%%%%%%%%%%%%%%%%%%%%%%%%%%%%%
%%%%%%%%%%%%%%%%%%%%%%%%%%%%%%%%%%%%%%%%%%%%%%%%%%%%%%%%%%%%%%%%%%%%%%%%%%%%%%%%%%%%%%%%%%%%%%%%%%%%%

%\section{Theoretical Model} %\label{sec:Theory}

\textit{Theoretical model}.--We used the Heisenberg--Langevin operator approach to analyze the biphoton generation in double-$\Lambda$ SFWM~\cite{Kolchin}. More details can be found in the Supplemental Material. The photon generation rate, given by $R=\frac{c}{L}\langle\hat{a}^\dagger\hat{a}\rangle$ and derived from the annihilation operator $\hat{a}$, leads to the following expressions:
\begin{align}
R_s
=&\int\frac{d\omega}{2\pi}
\left(
|B|^2
+\sum_{jk,j'k'}\int_0^Ldz\,
P_{jk}^*\mathcal{D}_{jk^\dagger,j'k'}P_{j'k'}
\right)
\nonumber\\
%-----------------------------------
\equiv&\int d\omega\widetilde{R}_s(\omega),
\\
%%%%%%%%%%%%%%%%%%%%%%%%%%%%%%%%%%%%%
R_{as}
=&\int\frac{d\omega}{2\pi}
\left(
|C|^2
+\sum_{jk,j'k'}\int_0^Ldz\,
Q_{jk}\mathcal{D}_{jk,j'k'^\dagger}Q_{j'k'}^*
\right)
\nonumber\\ 
%-----------------------------------
\equiv&\int d\omega\widetilde{R}_{as}(\omega),
\end{align}
where $\mathcal{D}_{jk^\dagger,j'k'}$ and $\mathcal{D}_{jk,j'k'^\dagger}$ are diffusion coefficients, while $\widetilde{R}_{s(as)}$ represents the spectrum of Stokes (anti-Stokes) photons. The total photon generation rates, $R_s$ and $R_{as}$, comprise two components: correlated photons governed by the FWM process with coefficients $B$ and $C$, and uncorrelated photons due to vacuum field fluctuations represented by an integral term with diffusion coefficients. The pairing ratio, $r_p$, denote the ratio of the correlated photons to the total generated photons. Under ideal conditions, $R_s$ and $R_{as}$ are nearly equal. However, in experiments, $R_{as}$ is slightly smaller due to phase mismatch and ground state decoherence. Therefore, the biphoton generation rate $R_{\rm B}$ is contingent on $R_{as}$

In the SFWM process within the atomic ensemble, a spontaneously emitted Stokes photon from one atom may interact with nearby atoms, triggering stimulated Raman scattering. Unlike spontaneous Raman scattering, the Stokes photon generated by stimulated Raman scattering shares the same direction as the incident Stokes photon, enhancing directionality. This collective enhancement effect establishes paired correlation with the anti-Stokes photon through the FWM process, reflected in coefficients $B$ and $C$. However, as SFWM relies on vacuum field fluctuations, the generated photons exhibit isotropic (uncorrelated) nature, represented by the integral term with diffusion coefficients in Eqs. (1) and (2). While the paired correlations of these biphotons can be established through the FWM process, it requires a sufficiently high OD within the atomic ensemble.

In biphoton systems, the normalized Glauber second-order cross-correlation function $g_{s\text{-}as}^{(2)}(\tau)$ is often used alongside the photon generation rate. This function is a crucial parameter for evaluating the temporal correlation between biphotons. The derived theoretical expression is as follows:
\begin{align}
g_{s\text{-}as}^{(2)}(\tau)=&
1+
\frac{1}{R_sR_{as}}
\left|
\int \frac{d\omega}{2\pi}
e^{-i\omega\tau}
\left(
\vphantom{\frac{A}{B}}
B^*D
\right.
\right.
\nonumber\\
%-----------------------------------
&+\sum_{jk,j'k'}\int_0^Ldz
\left.
\left.
\vphantom{\frac{A}{B}}
P_{jk}^*\mathcal{D}_{jk^\dagger,j'k'}Q_{j'k'}\right)
\right|^2.
\end{align}
The integral term on the right-hand side of Eq. (3) reveals the correlation of biphotons. This term is equivalent to the wavepacket of the anti-Stokes single photon, conditioned on the post-selection of a Stokes single photon. This correlation provides valuable information for evaluating the biphoton source. For instance, the peak signal-to-background ratio, denoted as $r_{\rm SB}$, is defined as the maximum value of $[g_{s\text{-}as}^{(2)}(\tau)-1]$. It serves as a standard metric for assessing the nonclassicality of a biphoton source. In the case of a classical field, the Cauchy--Schwarz inequality universally applies:
$
\left[g_{s\text{-}as}^{(2)}(\tau)\right]^2\left[g_{s\text{-}s}^{(2)}(0)g_{as\text{-}as}^{(2)}(0)\right]^{-1}\leq1.
$
The normalized autocorrelation functions of the Stokes and anti-Stokes fields can be derived as $g_{s\text{-}s}^{(2)}(\tau)=1+R_s^{-2}\left|\int d\omega\widetilde{R}_{s}e^{-i\omega\tau}\right|^2$ and $g_{as\text{-}as}^{(2)}(\tau)=1+R_{as}^{-2}\left|\int d\omega\widetilde{R}_{as}e^{-i\omega\tau}\right|^2$. These equations indicate that both the Stokes and anti-Stokes fields exhibit thermal states, with $g_{s\text{-}s}^{(2)}(0)=g_{as\text{-}as}^{(2)}(0)=2$. Nonclassical behavior is observed when $r_{\rm SB}>1$. Additional details and initial proofs of the thermal field distributions for both the Stokes and anti-Stokes fields can be found in the Supplemental Material.

%%%%%%%%%%%%%%%%%%%%%%%%%%%%%%%%%%%%%%%%%%%%%%%%%%%%%%%%%%%%%%%%%%%%%%%%%%%%%%%%%%%%%%%%%%%%%%%%%%%%%
%%%%%%%%%%%%%%%%%%%%%%%%%%%%%%%%%%%%%%%%%%%%%%%%%%%%%%%%%%%%%%%%%%%%%%%%%%%%%%%%%%%%%%%%%%%%%%%%%%%%%

%\section{Results and Discussion} %\label{sec:Results}

\textit{Biphoton bandwidth}.--Figure \ref{fig:Slow light regime} presents the experimental biphoton coincidence count rate $R_{\rm C}$ (or biphoton temporal wavepacket) for various coupling field conditions. $R_{\rm C}$ is calculated as $R_{s}R_{as}g_{s\text{-}as}^{(2)}(\tau)\Delta T + R_{\rm env}$ (refer to Supplemental Material), where $R_{\rm env}$ accounts for environmental background count rates, arising from laser leakage or SPCM dark counts. For data processing, we used a time bin of $\Delta T=1/R_s$ to tally Stokes photons, enabling the post-selection of a single Stokes photon. This ensures that the background and correlated regions of the coincidence count rate correspond to $R_{\rm B}+R_{\rm env}$ and $r_p$, respectively. The time bin for detected anti-Stokes photons, $\Delta\tau=6.4$ ns, aligns with the time interval between experimental data points in Fig. \ref{fig:Slow light regime}.

%%%%%%%%%%
\FigTwo
%%%%%%%%%%

In Figs. \ref{fig:Slow light regime}(a) and \ref{fig:Slow light regime}(b), we set the coupling Rabi frequencies $\Omega_c$ to 4$\Gamma$ and 1$\Gamma$, respectively. With the OD fixed at 15, both cases yielded a measured biphoton generation rate $R_{\rm B}$ of approximately $3.4\times10^5\text{s}^{-1}$. The delay time in Fig. \ref{fig:Slow light regime}(b) is significantly longer than that in \ref{fig:Slow light regime}(a). This delay arises from two intrinsic properties of the double-$\Lambda$ SFWM system: the damped Rabi oscillation with a period denoted as $\tau_{\rm R}=2\pi/\sqrt{|\Omega_c|^2-\Gamma^2/4}$, and the delay time attributed to the EIT effect denoted as $\tau_{\rm EIT}=\Gamma{\rm OD}/|\Omega_c|^2$~\cite{Du}. Both characteristic times are influenced by the coupling field. The damped Rabi oscillation periods in Figs. \ref{fig:Slow light regime}(a) and \ref{fig:Slow light regime}(b) are calculated as 42 and 192 ns, respectively. As $\Omega_c$ decreases, the EIT effect causes anti-Stokes photons to propagate slowly. The EIT delay times in Figs. \ref{fig:Slow light regime}(a) and \ref{fig:Slow light regime}(b) are 25 and 398 ns, respectively. The overall delay time is determined by the larger of $\tau_{\rm R}$ and $\tau_{\rm EIT}$, i.e., max$(\tau_{\rm R}, \tau_{\rm EIT})$. Consequently, the behavior of the biphoton wavepacket in Fig. \ref{fig:Slow light regime}(b), where $\tau_{\rm EIT}$ dominates, exhibits characteristics reminiscent of slow light, with the slow light effect noticeable in the trailing edge of the biphoton wavepacket. Conversely, in Fig. \ref{fig:Slow light regime}(a), where $\tau_{\rm R}$ surpasses $\tau_{\rm EIT}$, subtle oscillatory features are present within the biphoton wavepacket.

Figures \ref{fig:Slow light regime}(c) and \ref{fig:Slow light regime}(d) demonstrate how changes in coupling detuning $\Delta_c$ affect the biphoton bandwidth. All experimental parameters were consistent with those in Fig. \ref{fig:Slow light regime}(b), except for the $\Delta_c$. A shorter delay time in Fig. \ref{fig:Slow light regime}(c) is observed due to the introduction of $\Delta_c=1\Gamma$, which reduces the effective OD and shortens the EIT-induced delay. Conversely, with $\Delta_c=3\Gamma$ in Fig. \ref{fig:Slow light regime}(d), the tail lengthens again. This behavior is attributed to damped Rabi oscillations, where a larger $\Delta_c$ weakens the interaction between the coupling field and the atomic medium, requiring more time to convert spinwave excitations into anti-Stokes photons. The values of $R_{\rm B}$ in Figs. \ref{fig:Slow light regime}(c) and \ref{fig:Slow light regime}(d) are $3.4\times10^5\text{s}^{-1}$ and $3.1\times10^5\text{s}^{-1}$, respectively. This demonstrates that by detuning the coupling field, we can control the biphoton bandwidth without significantly reducing $R_{\rm B}$. Further discussions can be found in the Supplemental Material.

%%%%%%%%%%%%%%%%%%%%%%%%%%%%%%%%%%%%%%%%%%%%%%%%%%%%%%%%%%%%%%%%%%%%%%%%%%%%%%%%%%%%%%%%%%%%%%%%%%%%%
%%%%%%%%%%%%%%%%%%%%%%%%%%%%%%%%%%%%%%%%%%%%%%%%%%%%%%%%%%%%%%%%%%%%%%%%%%%%%%%%%%%%%%%%%%%%%%%%%%%%%

%\subsection{High-Purity Biphotons and Pairing Ratio}

%%%%%%%%%
\FigThree
%%%%%%%%%

\textit{High-purity biphotons}.--Figure \ref{fig:high purity biphoton}(a) illustrates high-purity biphoton generation achieved with parameters $\Omega_d=0.5\Gamma$, $\Omega_c=4\Gamma$, and ${\rm OD}=10$. The theoretical $R_{\rm B}$ is calculated as $5.0\times10^4\text{s}^{-1}$. In experiments, $R_{\rm B}$ was determined by subtracting the total measured background count rate, $R_{\rm tot}=1.6\times10^5\text{s}^{-1}$, from the environmental background count rate, $R_{\rm env}=1.1\times10^5\text{s}^{-1}$. This yielded an experimental $R_{\rm B}$ of approximately $5.0\times10^4\text{s}^{-1}$, in close agreement with the theoretical prediction. In Fig. \ref{fig:high purity biphoton}(b), we theoretically calculated $R_{\rm B}$ to be $1.9\times10^6\text{s}^{-1}$ at $\Omega_d=3\Gamma$. The experimentally observed $R_{\rm B}$, obtained from measurements of $R_{\rm tot}=2.0\times10^6\text{s}^{-1}$ and $R_{\rm env}=1.2\times10^5\text{s}^{-1}$, also closely matches theoretical prediction. As $\Omega_d$ increases, both $R_{\rm B}$ and $R_{\rm tot}$ rise significantly. However, this also leads to a notable decrease in $r_{\rm SB}$, as shown in Fig. \ref{fig:high purity biphoton}(c). At $\Omega_d=0.5\Gamma$, we observed an experimental $r_{\rm SB}=241$, surpassing the Cauchy--Schwarz criterion by a factor of approximately $1.5\times10^4$. If the $R_{\rm env}$ in our experiment could be completely eliminated, it would lead to a more pronounced violation of the Cauchy--Schwarz criterion, exceeding the normal level by a factor of $5.9\times10^4$.

Figure \ref{fig:high purity biphoton}(d) shows the variation of $R_{\rm B}$ and $r_p$ with different $\Omega_d$ values. At $\Omega_d=0.5\Gamma$ and $\Omega_d=3\Gamma$, the corresponding $r_p$ values are 0.63 and 0.59, respectively. These experimental $r_p$ values were determined based on the area under the correlated biphoton wavepacket. The $r_p$ obtained from the area and those obtained from Eqs. (1) and (2) are equivalent, as detailed in the Supplemental Material. In the SFWM process, atomic ensembles play a crucial role in collectively enhancing the correlation between the Stokes and anti-Stokes fields along the applied light direction. Therefore, with a fixed OD, while increasing $\Omega_d$ can boost $R_{\rm B}$, the limited density of atomic ensembles constrains their ability to produce correlated photon pairs, leading to a slight decrease in $r_p$.

%%%%%%%%%
\FigFour
%%%%%%%%%

%%%%%%%%%%%%%%%%%%%%%%%%%%%%%%%%%%%%%%%%%%%%%%%%%%%%%%%%%%%%%%%%%%%%%%%%%%%%%%%%%%%%%%%%%%%%%%%%%%%%%
%%%%%%%%%%%%%%%%%%%%%%%%%%%%%%%%%%%%%%%%%%%%%%%%%%%%%%%%%%%%%%%%%%%%%%%%%%%%%%%%%%%%%%%%%%%%%%%%%%%%%

%\subsection{Highly Paired Ultrabright Biphotons}

\textit{Highly correlated biphotons}.--Figure \ref{fig:pairing ratio}(a) showcases the generation of ultrabright biphotons using specific parameters: $\Omega_d=3\Gamma$, $\Omega_c=4\Gamma$, $\Delta_d=5\Gamma$, and ${\rm OD}=20$, resulting in a remarkable theoretical $R_{\rm B}$ of $1.3\times10^7\text{s}^{-1}$. This exceeds rates reported in the literature for the double-$\Lambda$ SFWM scheme. Under these conditions, the experimental total background count rate was $R_{\rm tot}=1.3\times10^7\text{s}^{-1}$, with environmental background at $R_{\rm env}=2.3\times10^5\text{s}^{-1}$, accounting for only 1.8$\%$ of the total. Thus, in this high $R_{\rm B}$ scenario, the primary source of background count arises from the high photon generation rate rather than environmental factors. Furthermore, in this scenario, the measured $r_{\rm SB}$ was 2.4, surpassing the Cauchy-Schwarz criterion by a factor of 2.9, while $r_p$ was only 0.61. Although increasing the coupling power can enhance $r_{\rm SB}$, as demonstrated in Fig. \ref{fig:Slow light regime}, it does not lead to corresponding improvements in $r_p$. To enhance both $r_p$ and $r_{\rm SB}$, we further increased the OD. In addition to ${\rm OD}=20$, we measured the biphoton wavepacket at ${\rm OD}=40$, $60$, $80$, $100$, and $120$. We fine-tuned $\Omega_c$ to maintain a consistent biphoton bandwidth, while keeping $\Omega_d=3\Gamma$ constant and adjusting $\Delta_d$ to maintain a theoretucal $R_{\rm B}$ of $1.3\times10^7 \text{s}^{-1}$. Specific parameters can be found in the Supplemental Material. 

In Fig. \ref{fig:pairing ratio}(b), we present the scenario with ${\rm OD}=120$, $\Omega_c=8.8\Gamma$, and $\Delta_d=14.9\Gamma$. Here, the measured $r_{\rm SB}$ at 4.2 exceeds the Cauchy--Schwarz criterion by 6.8 times. An evident positive correlation emerges between increased OD and enhanced $r_p$, resulting in a higher $r_{\rm SB}$ due to augmented coincidence receptions [Fig. \ref{fig:pairing ratio}(c)]. This enhancement stems from the increased accumulation of biphoton correlations along a specific direction at higher OD values. Photons generated at higher OD levels are more likely to encounter subsequent atoms, amplifying the collective enhancement through the FWM process. Furthermore, while $r_{\rm SB}$ can also be improved by increasing $\Omega_c$, this approach does not enhance $r_p$, and therefore cannot improve the generation rate of temporally correlated biphotons. Figure \ref{fig:pairing ratio}(d) illustrates the relationships between $R_{\rm B}$ and $r_p$ with OD. At ${\rm OD}=120$, we observed the highest $r_p$ of 0.89, indicating a significant improvement in correlated photon pair generation. The experimental $R_{\rm B}=1.3\times10^7\text{s}^{-1}$ signifies the successful generation of approximately $1.2\times10^7$ pairs of correlated photons per second. Additionally, the Fourier transform of $\left(R_{\rm C}-R_{\rm tot}\right)$ reveals a biphoton bandwidth of approximately 24 MHz, resulting in a spectral brightness of the biphoton source at $5.4\times10^5\text{s}^{-1}\text{MHz}^{-1}$, surpassing the highest achieved by sub-megahertz biphoton sources~\cite{Yu}. These results highlight the crucial role of high OD in SFWM-based biphoton sources, allowing for higher values of $R_{\rm B}$ and $r_p$. This enables the generation of a large quantity of high-quality correlated photon pairs for use in various quantum systems.

%%%%%%%%%%%%%%%%%%%%%%%%%%%%%%%%%%%%%%%%%%%%%%%%%%%%%%%%%%%%%%%%%%%%%%%%%%%%%%%%%%%%%%%%%%%%%%%%%%%%%
%%%%%%%%%%%%%%%%%%%%%%%%%%%%%%%%%%%%%%%%%%%%%%%%%%%%%%%%%%%%%%%%%%%%%%%%%%%%%%%%%%%%%%%%%%%%%%%%%%%%%

%\section{Conclusions} %\label{sec:Conclusion}

\textit{Conclusion}.--Our investigation into the biphoton pairing ratio, utilizing the double-$\Lambda$ SFWM in cold $^{87}$Rb atoms, revealed a marginal decrease with higher biphoton generation rates. Nonetheless, this trend can be effectively addressed by elevating the atomic ensemble density. The highest pairing ratio observed was 0.89 at an OD of 120, accompanied by an ultrabright biphoton generation rate of up to $1.3\times10^7\text{s}^{-1}$, surpassing previously reported rates achieved via the double-$\Lambda$ SFWM scheme. Furthermore, our experiment demonstrated the highest signal-to-background ratio of the biphoton wavepacket at 241, achieved at a low biphoton generation rate of $5.0\times10^4\text{s}^{-1}$. This outstanding performance exceeded the Cauchy--Schwarz criterion by approximately $1.5\times10^4$ times. These results underscore the capability of the double-$\Lambda$ SFWM scheme in advancing biphoton sources for future quantum technologies.

%%%%%%%%%%%%%%%%%%%%%%%%%%%%%%%%%%%%%%%%%%%%%%%%%%%%%%%%%%%%%%%%%%%%%%%%%%%%%%%%%%%%%%%%%%%%%%%%%%%%%
%%%%%%%%%%%%%%%%%%%%%%%%%%%%%%%%%%%%%%%%%%%%%%%%%%%%%%%%%%%%%%%%%%%%%%%%%%%%%%%%%%%%%%%%%%%%%%%%%%%%%

%\section*{ACKNOWLEDGEMENTS}

We thank Meng-Jung Lin, Chih-Min Yang, I-Chia Huang, and Ting-Ho Wu for their contributions to the initial setup of the experimental system. This work was supported by the National Science and Technology Council of Taiwan under Grant Nos. of 112-2112-M-006-034, 111-2639-M-007-001-ASP, and 111-2119-M-007-007.

\onecolumngrid
$\vphantom{1}$
\twocolumngrid

%%%%%%%%%%%%%%%%%%%%%%%%%%%%%%%%%%%%%%%%%%%%%%%%%%%%%%%%%%%%%%%%%%%%%%%%%%%%%%%%%%%%%%%%%%%%%%%%%%%%%
%%%%%%%%%%%%%%%%%%%%%%%%%%%%%%%%%%%%%%%%%%%%%%%%%%%%%%%%%%%%%%%%%%%%%%%%%%%%%%%%%%%%%%%%%%%%%%%%%%%%%

%%%%%%%%%%%%%%%%%%%%%%%%%%%%%%%%%%%%%%%%%%%%%%%%%%%%%%%%%%%%%%%%%%%%%%%%%%%%%%%%%%%%%%%%%%%%%%%%%%%%%
%%%%%%%%%%%%%%%%%%%%%%%%%%%%%%%%%%%%%%%%%%%%%%%%%%%%%%%%%%%%%%%%%%%%%%%%%%%%%%%%%%%%%%%%%%%%%%%%%%%%%

% Supplemental Material

\onecolumngrid
\newpage

\begin{center}
\fontsize{12}{14.5}\selectfont
\textbf{Supplemental Material for Observation of Highly Correlated Ultrabright Biphotons Through Increased Atomic Ensemble Density in Spontaneous Four-Wave Mixing}
%%%%%%%%%%%%%%%%%%%%%%%%%%%%%%%%%%%%%%%%%%%%%%%%
\\
\fontsize{13}{10}\selectfont$\vphantom{1}$
\\
\fontsize{10}{12.5}\selectfont
Jiun-Shiuan Shiu,$^{1,2}$ Zi-Yu Liu,$^{1,2}$ Chin-Yao Cheng,$^{1,2}$ Yu-Chiao Huang,$^{1,2}$ Ite A. Yu,$^{3,4}$ Ying-Cheng
\\
Chen,$^{5}$ Chih-Sung Chuu,$^{3,4}$ Che-Ming Li,$^{2,6}$ Shiang-Yu Wang,$^7$ and Yong-Fan Chen$^{1,2}$\\
%%%%%%%%%%%%%%%%%%%%%%%%%%%%%%%%%%%%%%%%%%%%%%%%
\fontsize{9.2}{10.5}\selectfont
\textit{
$^1$Department of Physics, National Cheng Kung University, Tainan 70101, Taiwan\\ 
	$^2$Center for Quantum Frontiers of Research $\&$ Technology, Tainan 70101, Taiwan\\ 
	$^3$Center for Quantum Science and Technology, National Tsing Hua University, Hsinchu 30013, Taiwan\\
	$^4$Department of Physics, National Tsing Hua University, Hsinchu 30013, Taiwan\\
	$^5$Institute of Atomic and Molecular Sciences, Academia Sinica, Taipei 10617, Taiwan\\
	$^6$Department of Engineering Science, National Cheng Kung University, Tainan 70101, Taiwan\\
	$^7$Institute of Astronomy and Astrophysics, Academia Sinica, Taipei 10617, Taiwan}
%%%%%%%%%%%%%%%%%%%%%%%%%%%%%%%%%%%%%%%%%%%%%%%%
\\
\fontsize{35}{10}\selectfont$\vphantom{1}$
\\
\begin{minipage}{0.785\textwidth}
\fontsize{9}{10.5}\selectfont
\hspace{1em}This supplemental material provides a comprehensive explanation of the theoretical model, delving into the details of experimental parameters and data analysis. In the first section, we introduce the theoretical model, covering the derivation of the field operator evolution in the double-$\Lambda$ spontaneous four-wave mixing atomic ensemble, biphoton generation rate, corresponding correlation functions, and the proof that Stokes and anti-Stokes photons adhere to a thermal state distribution. The second section elaborates on experimental measurements, including coincidence count rate, biphoton pairing ratio, and collection efficiency. Additionally, it discusses the temporal profile of the biphoton wavepacket and provides the experimental parameters presented in Figure 4 of the main text.
\end{minipage}
\end{center}

%%%%%%%%%%%%%%%%%%%%%%%%%%%%%%%%%%%%%%%%%%%%%%%%%%%%%%%%%%%%%%%%%%%%%%%%%%%%%%%%%%%%%%%%%%%%%%%%%%%%%
%%%%%%%%%%%%%%%%%%%%%%%%%%%%%%%%%%%%%%%%%%%%%%%%%%%%%%%%%%%%%%%%%%%%%%%%%%%%%%%%%%%%%%%%%%%%%%%%%%%%%

\maketitle

\newcommand{\FigSTwo}{
    \begin{figure}[b]
    \centering
    \includegraphics[width = 8.7 cm]{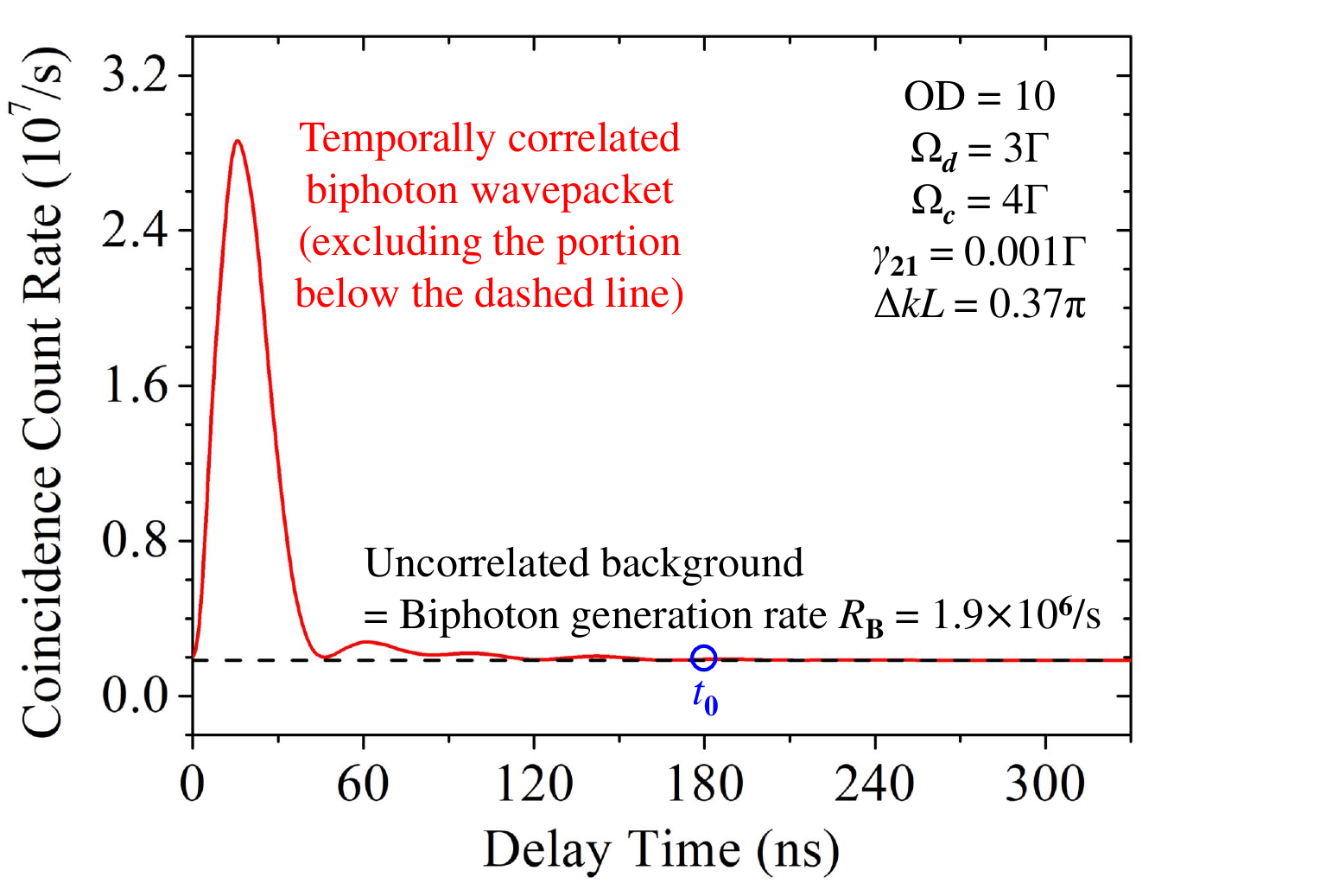}
    \caption{
The coincidence count rate corresponds to the parameters outlined in Fig. 3(b) of the main text. The red solid curve depicts the theoretically predicted coincidence count rate, while the black dashed line represents the uncorrelated background, which is equivalent to the biphoton generation rate.
}
    \label{fig:S2}
    \end{figure}
}

\newcommand{\FigSOne}{
    \begin{figure}[b]
    \centering
    \includegraphics[width = 18 cm]{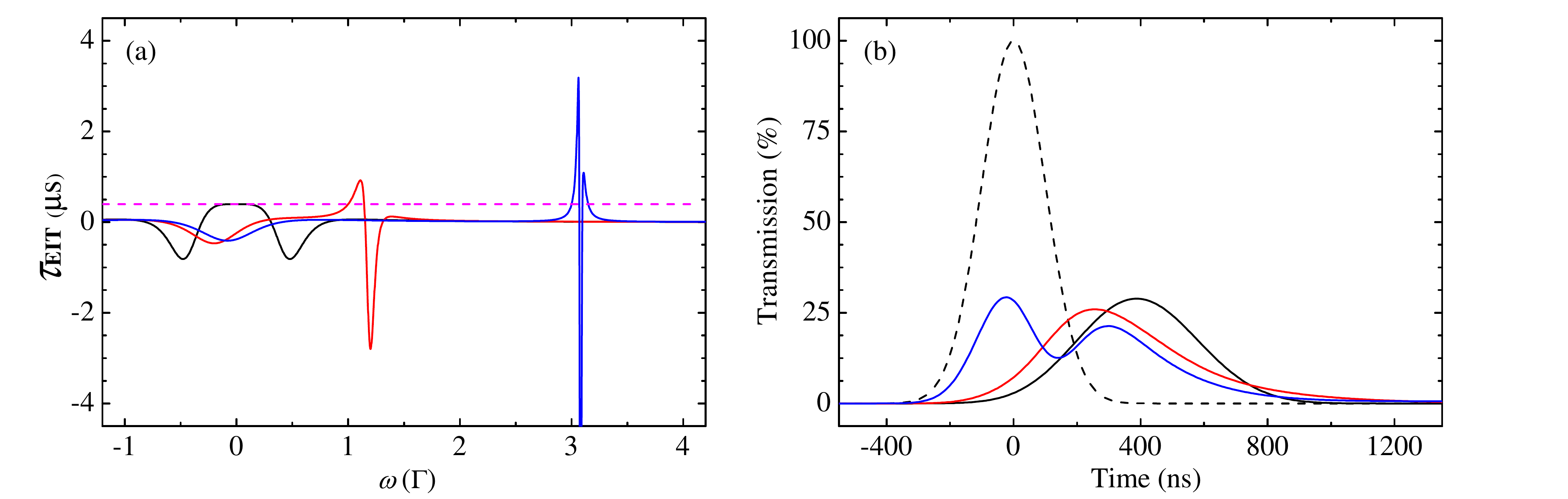}
    \caption{
EIT slow light effect. Parameters: ${\rm OD}=15$, $\Omega_c=1\Gamma$, $\gamma_{21}=0\Gamma$. (a) EIT delay time $\tau_{\rm EIT}$ as a function of detuning $\omega$. $\tau_{\rm EIT}$ is given by Eq. (S53). The solid black, red, and blue curves represent cases with $\Delta_c=0\Gamma$, $\Delta_c=1\Gamma$, and $\Delta_c=3\Gamma$, respectively. The magenta dashed line indicates a delay time of 398 ns calculated using $\tau_{\rm EIT}=\Gamma\,{\rm OD}/|\Omega_c|^2$. (b) Temporal profiles passing through the EIT medium. The black dashed curve represents the input Gaussian pulse with a $1/e^2$-full width of 400 ns. The solid black, red, and blue curves correspond to cases with $\Delta_c=0\Gamma$, $\Delta_c=1\Gamma$, and $\Delta_c=3\Gamma$, respectively.
}
    \label{fig:S1}
    \end{figure}
}

%%%%%%%%%%%%%%%%%%%%%%%%%%%%%%%%%%%%%%%%%%%%%%%%%%%%%%%%%%%%%%%%%%%%%%%%%%%%%%%%
%%%%%%%%%% Prefix a "S" to all equations, figures, tables and reset the counter %%%%%%%%%%
\setcounter{equation}{0}
\setcounter{figure}{0}
\setcounter{table}{0}
\setcounter{page}{1}
\thispagestyle{empty} % 第一頁不顯示頁碼
\makeatletter
\renewcommand{\theequation}{S\arabic{equation}}
\renewcommand{\thefigure}{S\arabic{figure}}
\renewcommand{\thetable}{S\arabic{table}}
\renewcommand{\bibnumfmt}[1]{[S#1]}
%\newcommand{\citenumfont}[1]{S#1}
%\newcommand{\mycitenumfont}[1]{S#1}
%\newcommand{\mycite}[1]{\hyperref[#1]{S\cite{#1}}}

%%%%%%%%%%%%%%%%%%%%%%%%%%%%%%%%%%%%%%%%%%%%%%%%%%%%%%%%%%%%%%%%%%%%%%%%%%%%%%%%%%%%%%%%%%%%%%%%%%%%%
%%%%%%%%%%%%%%%%%%%%%%%%%%%%%%%%%%%%%%%%%%%%%%%%%%%%%%%%%%%%%%%%%%%%%%%%%%%%%%%%%%%%%%%%%%%%%%%%%%%%%

\quad\\
\quad\\
\quad\\

\section{Theoretical Model} 

\subsection{Derivation of Field Operators}

In this research, we used the Heisenberg--Langevin operator approach to characterize the biphotons generated with the spontaneous four-wave mixing (SFWM) process~\cite{KolchinS}. The slowly varying collective atomic operator $\hat{\sigma}_{jk}(z,t)=\sum_{l=1}^{N_z}\hat{\sigma}^l_{jk}(z,t)/N_z$ was used to model the intricate atomic dynamics during the interaction, where $N_z$ is the number of atoms within the infinitesimal one-dimensional (1D) spatial interval $dz$. Here, $\hat{\sigma}^l_{jk}$ denotes the adiabatic atomic operator of the $l$th atom that has been transformed by the slowly varying amplitude. The dynamics of slowly varying collective atomic operators are governed by the Heisenberg--Langevin equations (HLEs): $\frac{\partial}{\partial t}\hat{\sigma}_{jk}=\frac{i}{\hbar}[\hat{H},\hat{\sigma}_{jk}]+\hat{R}_{jk}+\hat{F}_{jk}$, where $\hat{R}_{jk}$ and $\hat{F}_{jk}$ are the adiabatic relaxation term and the Langevin noise, respectively. The interaction Hamiltonian $\hat{H}$ of the double-$\Lambda$ SFWM atomic ensemble is
\begin{align}
\hat{H}=&-\frac{\hbar N}{2L}\int_0^Ldz
\left(
\Omega_d\hat{\sigma}_{31}
+\Omega_c\hat{\sigma}_{42}
+2g_s\hat{a}_s\hat{\sigma}_{32}\right.
\left.
+2g_{as}\hat{a}_{as}\hat{\sigma}_{41}
 e^{-i\Delta kz}
+\Delta_d\hat{\sigma}_{33}
+\Delta_c\hat{\sigma}_{44}
+{\rm H.c.}\right),
\end{align}
where $N$ and $L$ are the total number of atoms and the atomic medium length, respectively. The applied high-intensity coupling and driving fields, characterized by the semiclassical Rabi frequency $\Omega_{c(d)}$, induce the generation of Stokes and anti-Stokes photons, which are treated as quantum fields. Here, $g_s = d_{32}\sqrt{\bar{\omega}_s/2\hbar\epsilon_0V}$ is the coupling constant between the Stokes field and atoms, where $d_{32}$ indicates the dipole moment of the $|2\rangle\leftrightarrow|3\rangle$ transition, $\bar{\omega}_s$ is the center angular frequency of Stokes field, $\epsilon_0$ is the vacuum permittivity, and $V$ is the interaction volume. The coupling constant $g_{as}=d_{41}\sqrt{\bar{\omega}_{as}/2\hbar\epsilon_0V}$ between the anti-Stokes field and atoms is described in the same manner. The operator $\hat{a}_{s(as)}$ is the slowly varying annihilation operator of the Stokes (anti-Stokes) field. The phase shift $\Delta kz\approx(\vec{k}_d-\vec{k}_s+\vec{k}_c-\vec{k}_{as})\cdot\vec{z}$ represents the phase-mismatching value in the 1D approximation. All relevant HLEs for the double-$\Lambda$ SFWM system can be derived as follows: 
\begin{align}
	% 11 --------------------------------------------
	\frac{\partial\hat{\sigma}_{11}}{\partial t}=& i
	\left[
	\frac{\Omega_d^*}{2}\hat{\sigma}_{13}
	- \frac{\Omega_d}{2}\hat{\sigma}_{31}
	+ g_{as}^*\hat{a}_{as}^{\dagger}\hat{\sigma}_{14}
	e^{i\Delta kz}
	- g_{as}\hat{a}_{as}\hat{\sigma}_{41}e^{-i\Delta kz}
	\right] 
	+\Gamma_{31}\hat{\sigma}_{33}
	+\Gamma_{41}\hat{\sigma}_{44}
	+\hat{F}_{11},\\
	% 22 --------------------------------------------
	\frac{\partial\hat{\sigma}_{22}}{\partial t}=& i
	\left[
	\frac{\Omega_c^*}{2}\hat{\sigma}_{24}
	- \frac{\Omega_c}{2}\hat{\sigma}_{42}
	+ g_s^*\hat{a}_s^{\dagger}\hat{\sigma}_{23}
	- g_s\hat{a}_s\hat{\sigma}_{32}
	\right]
	+\Gamma_{32}\hat{\sigma}_{33}
	+\Gamma_{42}\hat{\sigma}_{44}
	+\hat{F}_{22},\\
	% 33 --------------------------------------------
	\frac{\partial\hat{\sigma}_{33}}{\partial t}=& i
	\left[
	\frac{\Omega_d}{2}\hat{\sigma}_{31}
	- \frac{\Omega_d^*}{2}\hat{\sigma}_{13}
	+ g_s\hat{a}_s\hat{\sigma}_{32}
	- g_s^*\hat{a}_s^{\dagger}\hat{\sigma}_{23}
	\right]
	-(\Gamma_{31}+\Gamma_{32})\hat{\sigma}_{33}
	+\hat{F}_{33},\\
	% 44 --------------------------------------------
	\frac{\partial\hat{\sigma}_{44}}{\partial t}=& i
	\left[
	\frac{\Omega_c}{2}\hat{\sigma}_{42}
	- \frac{\Omega_c^*}{2}\hat{\sigma}_{24}
	+ g_{as}\hat{a}_{as}\hat{\sigma}_{41}e^{-i\Delta kz}
	- g_{as}^*\hat{a}_{as}^{\dagger}\hat{\sigma}_{14}
	e^{i\Delta kz}
	\right]
	-(\Gamma_{41}+\Gamma_{42})\hat{\sigma}_{44}
	+\hat{F}_{44},\\
	% 13 --------------------------------------------
	\frac{\partial\hat{\sigma}_{13}}{\partial t}=& i
	\left[
	\frac{\Omega_d}{2}
	(\hat{\sigma}_{11}-\hat{\sigma}_{33})
	- g_{as}\hat{a}_{as}\hat{\sigma}_{43}e^{-i\Delta kz}
	+ g_s\hat{a}_s\hat{\sigma}_{12}
	\right]
	-\frac{\gamma_{31}-2i\Delta_d}{2}\hat{\sigma}_{13}
	+\hat{F}_{13},\\
	% 24 --------------------------------------------
	\frac{\partial\hat{\sigma}_{24}}{\partial t}=& i
	\left[
	\frac{\Omega_c}{2}
	(\hat{\sigma}_{22}-\hat{\sigma}_{44})
	- g_s\hat{a}_s\hat{\sigma}_{34}
	+ g_{as}\hat{a}_{as}\hat{\sigma}_{21}e^{-i\Delta kz}
	\right]
	-\frac{\gamma_{42}-2i\Delta_c}{2}\hat{\sigma}_{24}
	+\hat{F}_{24},\\
	% 21 --------------------------------------------
	\frac{\partial\hat{\sigma}_{21}}{\partial t}=& i
	\left[
	\frac{\Omega_d^*}{2}\hat{\sigma}_{23}
	- \frac{\Omega_c}{2}\hat{\sigma}_{41}
	- g_s\hat{a}_s\hat{\sigma}_{31}
	+ g_{as}^*\hat{a}_{as}^{\dagger}\hat{\sigma}_{24}e^{i\Delta kz}
	\right]
	-\frac{\gamma_{21}}{2}\hat{\sigma}_{21}
	+\hat{F}_{21},\\
	% 23 --------------------------------------------
	\frac{\partial\hat{\sigma}_{23}}{\partial t}=& i
	\left[
	g_s\hat{a}_s(\hat{\sigma}_{22}-\hat{\sigma}_{33})
	- \frac{\Omega_c}{2}\hat{\sigma}_{43}
	+ \frac{\Omega_d}{2}\hat{\sigma}_{21}
	\right]
	-\frac{\gamma_{32}-2i\Delta_d}{2}\hat{\sigma}_{23}
	+\hat{F}_{23},\\
	% 41 --------------------------------------------
	\frac{\partial\hat{\sigma}_{41}}{\partial t}=& i
	\left[
	- g_{as}^*\hat{a}_{as}^{\dagger}e^{i\Delta kz}
	(\hat{\sigma}_{11}-\hat{\sigma}_{44})
	+\frac{\Omega_d^*}{2}\hat{\sigma}_{43}
	- \frac{\Omega_c^*}{2}\hat{\sigma}_{21}
	\right]
	-\frac{\gamma_{41}+2i\Delta_c}{2}\hat{\sigma}_{41}
	+\hat{F}_{41},\\
	% 43 --------------------------------------------
	\frac{\partial\hat{\sigma}_{43}}{\partial t}=& i
	\left[
	\frac{\Omega_d}{2}\hat{\sigma}_{41}
	- \frac{\Omega_c^*}{2}\hat{\sigma}_{23}
	+ g_s\hat{a}_s\hat{\sigma}_{42}
	- g_{as}^*\hat{a}_{as}^{\dagger}\hat{\sigma}_{13}
	e^{i\Delta kz}
	\right]
	-\frac{\gamma_{43}-2i\Delta_d+2i\Delta_c}{2}\hat{\sigma}_{43}
	+\hat{F}_{43}.
\end{align}
Here $\Gamma_{jk}$ is the spontaneous decay rate from the excited state $|j\rangle$ to the ground state $|k\rangle$. $\gamma_{jk}$ is the decoherence rate corresponding to the states $|j\rangle$ and $|k\rangle$, which is due to the dephasing from the spontaneous emission. Of these, although both the states $|1\rangle$ and $|2\rangle$ are ground states, the term $\gamma_{21}$ is introduced because of some effects such as atomic collisions and magnetic inhomogeneity. In our system, $\Gamma_{31}=\Gamma_{32}=\Gamma_{41}=\Gamma_{42}=\frac{\Gamma}{2}$, where $\Gamma$ is the spontaneous decay rate of the rubidium-87 D$_2$ line, and therefore we have $\gamma_{31}=\gamma_{32}=\gamma_{41}=\gamma_{42}=\Gamma$ and $\gamma_{43}=2\Gamma$.

To simplify the solution of the HLEs, we employed perturbation theory. By neglecting the influence of the quantum field operator $\hat{a}_{s(as)}$ in the HLEs, we can derive the zeroth-order steady-state collective atomic operators in the form of $\hat{\sigma}_{jk}^{(0)}=\langle\hat{\sigma}_{jk}^{(0)}\rangle+\sum_{mn}\epsilon_{jk}^{mn}\hat{F}_{mn}$, with expectation values:
$\langle\hat{\sigma}_{11}^{(0)}\rangle=\frac{|\Omega_c|^2(\Gamma^2+4\Delta_d^2)+|\Omega_d|^2|\Omega_c|^2}{M}$, 
$\langle\hat{\sigma}_{22}^{(0)}\rangle=\frac{|\Omega_d|^2(\Gamma^2+4\Delta_c^2)+|\Omega_d|^2|\Omega_c|^2}{M}$, 
$\langle\hat{\sigma}_{33}^{(0)}\rangle=\langle\hat{\sigma}_{44}^{(0)}\rangle=\frac{|\Omega_d|^2|\Omega_c|^2}{M}$, 
$\langle\hat{\sigma}_{13}^{(0)}\rangle=\frac{i(\Gamma+2i\Delta_d)|\Omega_c|^2\Omega_d}{M}$, and 
$\langle\hat{\sigma}_{24}^{(0)}\rangle=\frac{i(\Gamma+2i\Delta_c)|\Omega_d|^2\Omega_c}{M}$. Here, $M=|\Omega_d|^2(\Gamma^2+4\Delta_c^2)+|\Omega_c|^2(\Gamma^2+4\Delta_d^2)+4|\Omega_d|^2|\Omega_c|^2$. Despite our assumption that both input fields operate in a single mode described by steady-state zeroth-order HLEs, the solutions of the zeroth-order collective atomic operators still exhibit time-varying terms due to the presence of the vacuum field. These time-varying terms lead to a dynamic equilibrium, which is reflected in the linear combination of $\hat{F}_{mn}$. By substituting the derived zeroth-order adiabatic collective atomic operators back into the HLEs, we can obtain the first-order HLEs as follows:
\begin{align}
&\frac{\partial\hat{\sigma}_{21}^{(1)}}{\partial t}=
i
\left[
  \frac{\Omega_d^*}{2}\hat{\sigma}_{23}^{(1)}
  -\frac{\Omega_c}{2}\hat{\sigma}_{41}^{(1)}
  -g_s\hat{a}_s\langle\hat{\sigma}_{31}^{(0)}\rangle
  +g_{as}^*\hat{a}_{as}^\dagger e^{i\Delta k z}\langle\hat{\sigma}_{24}^{(0)}\rangle
\right]
-\frac{\gamma_{21}}{2}\hat{\sigma}_{21}^{(1)}
+\hat{F}_{21},
\\
%%%%%%%%%%%%%%%%%%%%%%%%%%%%%%%%%%%%%%%%%%%%%%%%%%%%%%%%%%
%%%%%%%%%%%%%%%%%%%%%%%%%%%%%%%%%%%%%%%%%%%%%%%%%%%%%%%%%%
&\frac{\partial\hat{\sigma}_{23}^{(1)}}{\partial t}=
i
\left[
  g_s\hat{a}_s
    \left(\langle\hat{\sigma}_{22}^{(0)}\rangle
          -\langle\hat{\sigma}_{33}^{(0)}\rangle\right)
  -\frac{\Omega_c}{2}\hat{\sigma}_{43}^{(1)}
  +\frac{\Omega_d}{2}\hat{\sigma}_{21}^{(1)}
\right]
-\frac{\Gamma-2i\Delta_d}{2}\hat{\sigma}_{23}^{(1)}
+\hat{F}_{23},
\\
%%%%%%%%%%%%%%%%%%%%%%%%%%%%%%%%%%%%%%%%%%%%%%%%%%%%%%%%%%
%%%%%%%%%%%%%%%%%%%%%%%%%%%%%%%%%%%%%%%%%%%%%%%%%%%%%%%%%%
&\frac{\partial\hat{\sigma}_{41}^{(1)}}{\partial t}=
i
\left[
  -g_{as}^*\hat{a}_{as}^\dagger e^{i\Delta k z}
    \left(\langle\hat{\sigma}_{11}^{(0)}\rangle
          -\langle\hat{\sigma}_{44}^{(0)}\rangle\right)
  +\frac{\Omega_d^*}{2}\hat{\sigma}_{43}^{(1)}
  -\frac{\Omega_c^*}{2}\hat{\sigma}_{21}^{(1)}
\right]
-\frac{\Gamma+2i\Delta_c}{2}\hat{\sigma}_{41}^{(1)}
+\hat{F}_{41},
\\
%%%%%%%%%%%%%%%%%%%%%%%%%%%%%%%%%%%%%%%%%%%%%%%%%%%%%%%%%%
%%%%%%%%%%%%%%%%%%%%%%%%%%%%%%%%%%%%%%%%%%%%%%%%%%%%%%%%%%
&\frac{\partial\hat{\sigma}_{43}^{(1)}}{\partial t}=
i
\left[
  \frac{\Omega_d}{2}\hat{\sigma}_{41}^{(1)}
  -\frac{\Omega_c^*}{2}\hat{\sigma}_{23}^{(1)}
  +g_s\hat{a}_s\langle\hat{\sigma}_{42}^{(0)}\rangle
  -g_{as}^*\hat{a}_{as}^\dagger e^{i\Delta k z}\langle\hat{\sigma}_{13}^{(0)}\rangle
\right]
-(\Gamma-i\Delta_d+i\Delta_c)\hat{\sigma}_{43}^{(1)}
+\hat{F}_{43}.
\end{align}
When substituting the zeroth-order results into the first-order HLEs, we omitted the product of the field ladder operators and the Langevin noises $\hat{F}_{mn}$ due to their negligible effects. To investigate biphoton spectra, we applied the Fourier transform on Eqs. (S12)--(S15). We chose the angular frequency shift, represented as $\omega$, relative to the central angular frequency of the Stokes field, $\bar{\omega}_s$, as the frequency-domain variable in the Fourier transform. In other words, $\widetilde{a}_s(z,\omega)=\frac{1}{2\pi}\int dt\hat{a}_s(z,t)e^{i\omega t}\equiv\mathcal{F}\lbrace\hat{a}_s(z,t)\rbrace$. To maintain the factor of the integrand as $e^{i\omega t}$, the Fourier transform of the creation operator should be adjusted to $\widetilde{a}_s^\dagger(z,-\omega)=\frac{1}{2\pi}\int dt\hat{a}_s^\dagger(z,t)e^{i\omega t}$. Similarly, we have $\widetilde{a}_{as}(z,\omega)=\mathcal{F}\lbrace\hat{a}_{as}(z,t)\rbrace$, $\widetilde{a}_{as}^\dagger(z,-\omega)=\mathcal{F}\lbrace\hat{a}_{as}^\dagger(z,t)\rbrace$, $\widetilde{\sigma}_{jk}^{(1)}(z,\omega)=\mathcal{F}\lbrace\hat{\sigma}_{jk}^{(1)}(z,t)\rbrace$, and $\widetilde{F}_{jk}(z,\omega)=\mathcal{F}\lbrace\hat{F}_{jk}(z,t)\rbrace$. Therefore, the solutions for the frequency-domain collective atomic operators $\widetilde{\sigma}_{23}$ and $\widetilde{\sigma}_{41}$ can be expressed as $\widetilde{\sigma}_{23(41)}=\epsilon_{23(41)}g_s\widetilde{a}_s+\eta_{23(41)}g_{as}^*\widetilde{a}_{as}^\dagger e^{i\Delta k z}+\sum_{jk}\xi_{jk}^{s(as)}\widetilde{F}_{jk}$, where $\lbrace jk\rbrace=\lbrace21,23,41,43\rbrace$. We combine these results with the frequency-domain Maxwell--Schr\"{o}dinger equations used to describe our experimental backward double-$\Lambda$ SFWM system as follows:
\begin{align}
&\left(
-\frac{i\omega}{c}+
\frac{\partial}{\partial z}
\right)\widetilde{a}_s
=\frac{iNg_s^*}{c}\widetilde{\sigma}_{23}^{(1)},
\\
%%%%%%%%%%%%%%%%%%%%%%%%%%%%%%%%%%%%%%%%%%%%%%%%%%%%%%%%%%
%%%%%%%%%%%%%%%%%%%%%%%%%%%%%%%%%%%%%%%%%%%%%%%%%%%%%%%%%%
&\left(-\frac{i\omega}{c}-
\frac{\partial}{\partial z}
\right)\widetilde{a}_{as}^\dagger
=-\frac{iNg_{as}}{c}\widetilde{\sigma}_{41}^{(1)}e^{-i\Delta k z}.
\end{align} 
We can simplify the expressions by introducing a change of variables, defining $\widetilde{a}_{as,\Delta k}^\dagger\equiv\widetilde{a}_{as}^\dagger e^{i\Delta kz}$ to remove the exponential phase variation. This allows us to reorganize Eqs. (S16) and (S17) into the following matrix form:
\begin{align}
\frac{\partial}{\partial z}
\begin{pmatrix}
\widetilde{a}_s \\ \widetilde{a}_{as,\Delta k}^\dagger
\end{pmatrix}
=
\begin{pmatrix}
g_R & \kappa_s \\ \kappa_{as} & \Gamma_{as}
\end{pmatrix}
\begin{pmatrix}
\widetilde{a}_s \\ \widetilde{a}_{as,\Delta k}^\dagger
\end{pmatrix}
+\frac{iN}{c}
\sum_{jk}
\begin{pmatrix}
g_s^*\xi_{jk}^s \\ g_{as}\xi_{jk}^{as}
\end{pmatrix}\widetilde{F}_{jk}.
\end{align}
The Raman gain coefficient, denoted as $g_R$, is expressed as: $g_R=\frac{iN|g_s|^2}{c}\epsilon_{23}+\frac{i\omega}{c}$. The first component, $\frac{iN|g_s|^2}{c}\epsilon_{23}$, characterizes the Raman process, while the second component, $\frac{i\omega}{c}$, represents the phase term arising from free evolution. The electromagnetically induced transparency (EIT) profile coefficient, denoted as $\Gamma_{as}$, takes the form: $\Gamma_{as}=\frac{iN|g_s|^2}{c}\eta_{41}+i\Delta k-\frac{i\omega}{c}$. In this expression, the first component, $\frac{iN|g_s|^2}{c}\eta_{41}$, describes the EIT characteristics, and the second term, $i\Delta k$, represents the phase-mismatching effect. Furthermore, we have the Stokes (anti-Stokes) coupling coefficients $\kappa_s$ and $\kappa_{as}$, defined as: $\kappa_s=\frac{iNg_s^*g_{as}^*}{c}\eta_{23}$ and $\kappa_{as} = \frac{iNg_sg_{as}}{c}\epsilon_{41}$. The coefficient $\kappa_s$ ($\kappa_{as}$) quantifies how the anti-Stokes (Stokes) field affects the Stokes (anti-Stokes) field.

The solution to Eq. (S18) can be obtained by directly integrating from $z=0$ to $z=L$, as illustrated below:
\begin{align}
\begin{pmatrix}
\widetilde{a}_s(L) \\ \widetilde{a}_{as,\Delta k}^\dagger(L)
\end{pmatrix}
=&
{\rm exp}\left[
\begin{pmatrix}
g_R & \kappa_s \\ \kappa_{as} & \Gamma_{as}
\end{pmatrix}L
\right]
\begin{pmatrix}
\widetilde{a}_s(0) \\ \widetilde{a}_{as,\Delta k}^\dagger(0)
\end{pmatrix}
+\frac{iN}{c}
\sum_{jk}
\int_0^Ldz\,
{\rm exp}\left[
\begin{pmatrix}
g_R & \kappa_s \\ \kappa_{as} & \Gamma_{as}
\end{pmatrix}(L-z)
\right]
\begin{pmatrix}
g_s^*\xi_{jk}^s \\ g_{as}\xi_{jk}^{as}
\end{pmatrix}\widetilde{F}_{jk}
\nonumber\\%------------------------
\equiv&
\begin{pmatrix}
A' & B' \\ C' & D'
\end{pmatrix}
\begin{pmatrix}
\widetilde{a}_s(0) \\ \widetilde{a}_{as,\Delta k}^\dagger(0)
\end{pmatrix}
+\sqrt{\frac{N}{c}}
\sum_{jk}
\int_0^Ldz\,
\begin{pmatrix}
P'_{jk} \\ Q'_{jk}
\end{pmatrix}\widetilde{F}_{jk}.
\end{align}
Since the Stokes and anti-Stokes fields are co-propagating outputs in the backward double-$\Lambda$ SFWM system, we now reorganize Eq. (S19) to describe this biphoton output as follows:
\begin{align}
\begin{pmatrix}
\widetilde{a}_s(L) \\ \widetilde{a}_{as,\Delta k}^\dagger(0)
\end{pmatrix}
=&
\begin{pmatrix}
A'-\frac{B'C'}{D'} & \frac{B'}{D'} \\
-\frac{C'}{D'} & \frac{1}{D'}
\end{pmatrix}
\begin{pmatrix}
\widetilde{a}_s(0) \\ \widetilde{a}_{as,\Delta k}^\dagger(L)
\end{pmatrix}
+
\sqrt{\frac{N}{c}}
\sum_{jk}
\int_0^Ldz\,
\begin{pmatrix}
P'_{jk}-\frac{B'}{D'}Q'_{jk} \\ -\frac{1}{D'}Q'_{jk}
\end{pmatrix}\widetilde{F}_{jk}
\nonumber\\%-------------------------
\equiv&
\begin{pmatrix}
A & B \\ C & D
\end{pmatrix}
\begin{pmatrix}
\widetilde{a}_s(0) \\ \widetilde{a}_{as,\Delta k}^\dagger(L)
\end{pmatrix}
+
\sqrt{\frac{N}{c}}
\sum_{jk}
\int_0^Ldz\,
\begin{pmatrix}
P_{jk} \\ Q_{jk}
\end{pmatrix}\widetilde{F}_{jk}.
\end{align}
$A$ and $D$ represent coefficients for mode preservation, elucidating the influence of the Stokes and anti-Stokes inputs on their respective outputs. On the other hand, $B$ and $C$ represent coefficients for mode conversion, describing the energy conversion between the Stokes and anti-Stokes fields, which is the process of photon mutual conversion in the double-$\Lambda$ four-wave mixing (FWM)~\cite{QFCS, HOMS}.

%%%%%%%%%%%%%%%%%%%%%%%%%%%%%%%%%%%%%%%%%%%%%%%%%%%%%%%%%%%%%%%%%%%%%%%%%%%%%%%%%%%%%%%%%%%%%%%%%%%%%
%%%%%%%%%%%%%%%%%%%%%%%%%%%%%%%%%%%%%%%%%%%%%%%%%%%%%%%%%%%%%%%%%%%%%%%%%%%%%%%%%%%%%%%%%%%%%%%%%%%%%

\subsection{Biphoton Generation Rate and Correlation Functions}

The biphoton generation rate is related to the photon flux passing through the interaction cross-section area with the atoms. The photon flux can be obtained using the Poynting vector, denoted as $\vec{S}=\frac{1}{\mu_0}\vec{E}\times\vec{B}^*$, which represents the energy flux per unit area of an electromagnetic wave. By dividing the surface integral of the Poynting vector by the energy of a single photon, and taking into account the quantized electric field in the form of $\vec{E}=\hat{e}
\sqrt{\frac{\hbar\bar{\omega}}{2\epsilon_0V}}
\hat{a}e^{i(\vec{k}\cdot\vec{r}-\omega t)}+{\rm H.c.}$, we can determine the quantum photon generation rate as follows:
\begin{align}
R
=\frac{1}{\hbar\bar{\omega}}\int\vec{S}\cdot d\vec{\sigma}
=\frac{c}{L}\langle\hat{a}^{\dagger}\hat{a}\rangle+\frac{c}{2L}.
\end{align}
The parameter $\hat{e}$ is the unit vector of the polarized direction. The term $\frac{c}{2L}$ on the right-hand side of Eq. (S21) arises from the zero-point energy of the vacuum field and does not contribute to the actual count of generated photons. Thus, the photon generation rate can be expressed as $R=\frac{c}{L}\langle\hat{a}^{\dagger}\hat{a}\rangle$. By employing the frequency-domain commutation relation $[\widetilde{a}(\omega'),\widetilde{a}^\dagger(-\omega)]=\frac{L}{2\pi c}\delta(\omega+\omega')$ and Einstein relation $\langle \widetilde{F}_{jk}^\dagger(z,-\omega)\widetilde{F}_{j'k'}(z',\omega')\rangle=\frac{L}{2\pi N}\mathcal{D}_{jk^\dagger,j'k'}\delta(z-z')\delta(\omega+\omega')$, we can derive the photon generation rates for the Stokes ($R_s$) and anti-Stokes ($R_{as}$) fields, as given by Eqs. (1) and (2) presented in the main text. In our SFWM system, it is worth noting that $g_p = g_s\equiv g$. To conform to commonly used experimental parameters, we perform the substitution $\frac{g^2N}{c}=\frac{\rm{OD}\Gamma}{4L}$, where OD denotes the optical depth of the atomic medium. In addition to the photon generation rate, we can also derive the normalized Glauber second-order cross-correlation function, denoted as $g_{s\text{-}as}^{(2)}(\tau)$, as depicted in Eq. (3) in the main text. Similarly, we can obtain the normalized autocorrelation functions $g_{s\text{-}s}^{(2)}(\tau)$ and $g_{as\text{-}as}^{(2)}(\tau)$ for the generated Stokes and anti-Stokes photons.
%%%%%%%%%%%%%%%%%%%%%%%%%%%%%%%%%%%%%%%%%%%%%%%%%%%%%%%%%%%%%%%%%%%%%%%%%%%%%%%%%%%%%%%%%%%%%%%%%%%%%
%%%%%%%%%%%%%%%%%%%%%%%%%%%%%%%%%%%%%%%%%%%%%%%%%%%%%%%%%%%%%%%%%%%%%%%%%%%%%%%%%%%%%%%%%%%%%%%%%%%%%

\subsection{Diffusion Coefficients and Einstein Relation}

The HLE for the $l$th atom is given by $\frac{\partial}{\partial t}\hat{\sigma}_{jk}^l=\frac{i}{\hbar}[\hat{H}^l,\hat{\sigma}_{jk}^l]+\hat{R}_{jk}^l+\hat{F}_{jk}^l$. According to the Markovian approximation, the correlation between Langevin noises follows a Dirac delta function as $\langle\hat{F}_{jk}^l(t)\hat{F}_{j'k'}^l(t')\rangle=\mathcal{D}_{jk,j'k'}\delta(t-t')$, where $\mathcal{D}_{jk,j'k'}$ represents the diffusion coefficient satisfying the Einstein relation~\cite{QOS}. In this case, the Einstein relation can be derived from the following equation:
\begin{align}
\frac{\partial}{\partial t}
\langle\hat{\sigma}_{jk}^l\hat{\sigma}_{j'k'}^l\rangle
=&
\left\langle
\left(
\frac{i}{\hbar}[\hat{H}^l,\hat{\sigma}_{jk}^l]+\hat{R}_{jk}^l+\hat{F}_{jk}^l
\right)
\hat{\sigma}_{j'k'}^l
\right\rangle
+
\left\langle
\hat{\sigma}_{jk}^l
\left(
\frac{i}{\hbar}[\hat{H}^l,\hat{\sigma}_{j'k'}^l]+\hat{R}_{j'k'}^l+\hat{F}_{j'k'}^l
\right)
\right\rangle
\nonumber\\%---------------------------
=&
\frac{i}{\hbar}
\langle
[\hat{H}^l,\hat{\sigma}_{jk}^l\hat{\sigma}_{j'k'}^l]
\rangle
+\langle\hat{R}_{jk}^l\hat{\sigma}_{j'k'}^l\rangle
+\langle\hat{\sigma}_{jk}^l\hat{R}_{j'k'}^l\rangle
+\langle\hat{F}_{jk}^l\hat{\sigma}_{j'k'}^l\rangle
+\langle\hat{\sigma}_{jk}^l\hat{F}_{j'k'}^l\rangle,
\end{align}
where the term of $\langle\hat{F}_{jk}^l\hat{\sigma}_{j'k'}^l\rangle$ can be expressed as 
\begin{align}
\langle\hat{F}_{jk}^l(t)\hat{\sigma}_{j'k'}^l(t)\rangle
=&
\langle\hat{F}_{jk}^l(t)\hat{\sigma}_{j'k'}^l(t-\Delta t)\rangle
+
\int_{t-\Delta t}^tdt'
\left\langle
\hat{F}_{jk}^l(t)\frac{d\hat{\sigma}_{j'k'}^l(t')}{dt'}
\right\rangle
\nonumber\\%---------------------------
=&
\frac{i}{\hbar}
\int_{t-\Delta t}^tdt'
\langle
\hat{F}_{jk}^l(t)[\hat{H}^l(t'),\hat{\sigma}_{j'k'}^l(t')]
\rangle
+
\int_{t-\Delta t}^tdt'
\left\langle
\hat{F}_{jk}^l(t)\hat{R}_{j'k'}^l(t')
\right\rangle
+
\int_{t-\Delta t}^tdt'
\left\langle
\hat{F}_{jk}^l(t)\hat{F}_{j'k'}^l(t')
\right\rangle.
\end{align}
The term $\langle\hat{F}_{jk}^l(t)\hat{\sigma}_{j'k'}^l(t-\Delta t)\rangle$ yields zero as the Langevin noises at time $t$ are incapable of influencing the atom at time $t-\Delta t$, owing to the nonanticipatory property. Additionally, given that the terms $[\hat{H}^l(t'),\hat{\sigma}_{j'k'}^l(t')]$ and $\hat{R}_{j'k'}^l(t')$ evolve at a slower rate compared to the Langevin noises, the integrals of the first two terms in Eq. (S23) both approach to zero. Consequently, Eq. (S23) simplifies to $\langle\hat{F}_{jk}^l\hat{\sigma}_{j'k'}^l\rangle=\mathcal{D}_{jk,j'k'}/2$. A similar derivation leads to the relation $\langle\hat{\sigma}_{jk}^l\hat{F}_{j'k'}^l\rangle=\mathcal{D}_{jk,j'k'}/2$. Employing these relations, Eq. (S22) can be further expressed as follows:
\begin{align}
\mathcal{D}_{jk,j'k'}=&
\frac{\partial}{\partial t}
\langle\hat{\sigma}_{jk}^l\hat{\sigma}_{j'k'}^l\rangle
-
\frac{i}{\hbar}
\langle
[\hat{H}^l,\hat{\sigma}_{jk}^l\hat{\sigma}_{j'k'}^l]
\rangle
-\langle\hat{R}_{jk}^l\hat{\sigma}_{j'k'}^l\rangle
-\langle\hat{\sigma}_{jk}^l\hat{R}_{j'k'}^l\rangle
\nonumber\\%---------------------------
=&
\delta_{kj'}\langle\hat{R}_{jk'}^l\rangle
-\langle\hat{R}_{jk}^l\hat{\sigma}_{j'k'}^l\rangle
-\langle\hat{\sigma}_{jk}^l\hat{R}_{j'k'}^l\rangle,
\end{align}
and this is commonly referred to as the Einstein relation. After establishing these relationships, we can extend our findings from the single-atom scenario to collective situations. In the context of an atomic ensemble system, we define the collective Langevin noise as $\hat{F}_{jk}=\sum_{l=1}^{N_z}\hat{F}_{jk}^l/N_z$, where $N_z=N {dz}/L$. With this definition in place, we proceed to formulate
\begin{align}
\langle
\hat{F}_{jk}(z,t)\hat{F}_{j'k'}(z',t')
\rangle
=
\frac{1}{N_z^2}\sum_{l,l'}
\langle
\hat{F}_{jk}^l\hat{F}_{j'k'}^{l'}
\rangle\delta_{ll'}\delta_{zz'}
=
\frac{1}{N_z^2}\sum_{l}
\langle
\hat{F}_{jk}^l\hat{F}_{j'k'}^{l}
\rangle\delta_{zz'}
=
\frac{\delta_{zz'}}{N_z}
\mathcal{D}_{jk,j'k'}\delta(t-t').
\end{align}
By using the relation $\delta_{zz'}=\delta(z-z'){dz}$, we can obtain the Einstein relations in both the time and frequency domains for the atomic ensemble system as follows:
\begin{align}
\langle
\hat{F}_{jk}(z,t)\hat{F}_{j'k'}(z',t')
\rangle=\frac{L}{N}\mathcal{D}_{jk,j'k'}\delta(z-z')\delta(t-t'). 
\end{align}
\begin{align}
\langle
\widetilde{F}_{jk}(z,\omega)\widetilde{F}_{j'k'}(z',\omega')
\rangle=\frac{L}{2\pi N}\mathcal{D}_{jk,j'k'}\delta(z-z')\delta(\omega+\omega').
\end{align}

%%%%%%%%%%%%%%%%%%%%%%%%%%%%%%%%%%%%%%%%%%%%%%%%%%%%%%%%%%%%%%%%%%%%%%%%%%%%%%%%%%%%%%%%%%%%%%%%%%%%%
%%%%%%%%%%%%%%%%%%%%%%%%%%%%%%%%%%%%%%%%%%%%%%%%%%%%%%%%%%%%%%%%%%%%%%%%%%%%%%%%%%%%%%%%%%%%%%%%%%%%%

\subsection{Proof of Thermal States for Stokes and Anti-Stokes Fields}

We employ the reduced density operator approach to derive the density matrices of the Stokes and anti-Stokes fields generated from the double-$\Lambda$ SFWM system. This method transforms the obtained field operator into the form of a density matrix~\cite{ScullyQOS, densityS, QM3S}. Consider an open system with a Stokes field $\rho^s(t)$, anti-Stokes field $\rho^{as}(t)$, and vacuum reservoir $\rho^R(t)$. Initially, these fields are assumed to be uncorrelated, meaning $\rho(0)=\rho^s(0)\otimes\rho^{as}(0)\otimes\rho^R(0)$. The dynamics of this open system can be described by $\rho(t)=U(t)\rho(0)U^\dagger(t)$, where $U(t)$ is the time evolution operator of the total density matrix. The Stokes field $\rho^{s}(t)$ can be obtained by tracing over the degree of freedom of $\rho^{as}(t)$ and $\rho^R(t)$ from the total density matrix, i.e. $\rho^{s}(t)={\rm Tr}_{as,R}\lbrace\rho(t)\rbrace$. The density matrix element of the generated Stokes field in the Fock-state basis can be given by
\begin{align}
\rho^s_{mn}(t)&=
%-------------------------------
\langle m|
{\rm Tr}_{as,R}\left\lbrace U(t)\rho(0)U^\dagger(t)\right\rbrace
|n\rangle
\nonumber\\&=
%-------------------------------
{\rm Tr}_s
\left\lbrace
\vphantom{\frac{A}{B}}
|n_s\rangle\langle m_s|\otimes
{\rm Tr}_{as,R}\left\lbrace U(t)\rho(0)U^\dagger(t)\right\rbrace
\right\rbrace
\nonumber\\&=
%-------------------------------
{\rm Tr}
\left\lbrace
\vphantom{\frac{A}{B}}
\left(
\vphantom{A^{B^B}_{C_C}}
|n_s\rangle\langle m_s|\otimes I_{as}\otimes I_R
\right)
U(t)\rho(0)U^\dagger(t)
\right\rbrace
\nonumber\\&=
%-------------------------------
{\rm Tr}
\left\lbrace
\vphantom{\frac{A}{B}}
U^\dagger(t)
\left(
\vphantom{A^{B^B}_{C_C}}
|n_s\rangle\langle m_s|\otimes I_{as}\otimes I_R
\right)
U(t)\rho(0)
\right\rbrace
\equiv
%-------------------------------
{\rm Tr}
\left\lbrace
\vphantom{\frac{A}{B}}
\hat{\rho}^{s}_{mn}(t)\rho(0)
\right\rbrace.
\end{align}
Consider the vacuum input $\rho(0)=|0_s\rangle\langle0_s|\otimes |0_{as}\rangle\langle0_{as}|\otimes \rho^R(0)$ in the double-$\Lambda$ SFWM system. By rewriting the vacuum density matrix with the annihilation and creation operators $|0\rangle\langle0|=\sum_{l=0}^\infty\frac{(-1)^l}{l!}[\hat{a}^\dagger(0)]^l[\hat{a}(0)]^l$~\cite{LouisellQOS}, we can derive the density matrix element of the Stokes field as follows:
\begin{align}
&
\rho^s_{mn}(t)
=
%-------------------------------
\frac{1}{\sqrt{m!n!}}\sum_{l=0}^\infty\frac{(-1)^l}{l!}
\langle
[\hat{a}_s^\dagger(t)]^{n+l}[\hat{a}_s(t)]^{m+l}
\rangle.
\end{align}
Using the inverse Fourier transform and combining Eq. (S20), the Stokes field operator in the time domain can be obtained as follows:
\begin{align}
\hat{a}_s(L,t)=&\int d\omega e^{-i\omega t}
\left[ 
A\widetilde{a}_s(0,\omega)
+B\widetilde{a}_{as}^{\dagger}(L,-\omega)
+\widetilde{F}_s(\omega)
\right],
\end{align}
where $\widetilde{F}_s(\omega)$ is a noise-correlated term that is a linear combination of $\widetilde{F}_{21}$, $\widetilde{F}_{23}$, $\widetilde{F}_{41}$, and $\widetilde{F}_{43}$. After substituting Eq. (S30) into Eq. (S29) and considering the initial vacuum input $\rho(0)$, we find that only the condition where $m=n$ contributes to a non-zero value of the density matrix element of the Stokes field. With these results, we can derive the density matrix element of the Stokes field as follows:
\begin{align}
&\rho^s_{nn}
=
\left\langle
 \sum_{l=0}^{\infty}\frac{(-1)^l}{l!n!}
%----------------------
\left\lbrace\int d\omega e^{-i\omega t}
 \left[
  B^*\widetilde{a}_{as}(L,\omega)
  +\widetilde{F}_s^{\dagger}(-\omega)
 \right]
\right\rbrace^{l+n}
\left\lbrace\int d\omega e^{-i\omega t}
 \left[
  B\widetilde{a}_{as}^{\dagger}(L,-\omega)
  +\widetilde{F}_s(\omega)
 \right]
\right\rbrace^{l+n}
\right\rangle.
\end{align}
To further simplify Eq. (S31), we define
$
\hat{\zeta}\equiv
\int d\omega e^{-i\omega t}
  B^*\widetilde{a}_{as}(L,\omega)
$ and $
\hat{\eta}\equiv
\int d\omega e^{-i\omega t}
\widetilde{F}_s(\omega)$, 
resulting in the following expression:
\begin{align}
\rho^s_{nn}
&=
\left\langle
 \sum_{l=0}^{\infty}\frac{(-1)^l}{l!n!}
 \left(\hat{\zeta}+\hat{\eta}^\dagger\right)^{l+n}
 \left(\hat{\zeta}^\dagger+\hat{\eta}\right)^{l+n}
\right\rangle\nonumber\\
%--------------------------------
&=
\left\langle
 \sum_{l=0}^{\infty}\frac{(-1)^l}{l!n!}
 \sum_{\alpha=0}^{l+n}C^{l+n}_\alpha
   \hat{\zeta}^\alpha(\hat{\eta}^\dagger)^{l+n-\alpha}
 \sum_{\beta=0}^{l+n}C^{l+n}_\beta
   (\hat{\zeta}^\dagger)^\beta\hat{\eta}^{l+n-\beta}
\right\rangle\nonumber\\
%--------------------------------
&=
\sum_{l=0}^{\infty}\frac{(-1)^l}{l!n!}
\sum_{\alpha=0}^{l+n}
\sum_{\beta=0}^{l+n}
C^{s+n}_\alpha C^{s+n}_\beta
 \left\langle
   \hat{\zeta}^\alpha(\hat{\zeta}^\dagger)^\beta
 \right\rangle
 \left\langle
   (\hat{\eta}^\dagger)^{l+n-\alpha}\hat{\eta}^{l+n-\beta}
 \right\rangle,
\end{align}
where we apply the binomial coefficient $C^a_b=\frac{a!}{b!(a-b)!}$, and only when $\beta$ equals $\alpha$, there is a non-zero contribution, leading to the following expression:
\begin{align}
\rho^s_{nn}
=
\sum_{l=0}^{\infty}\frac{(-1)^l}{l!n!}
\sum_{\alpha=0}^{l+n}
(C^{l+n}_\alpha)^2
 \left\langle
   \hat{\zeta}^\alpha(\hat{\zeta}^\dagger)^\alpha
 \right\rangle
 \left\langle
   (\hat{\eta}^\dagger)^{l+n-\alpha}\hat{\eta}^{l+n-\alpha}
 \right\rangle.
\end{align}
The operator product term $\left\langle\hat{\zeta}^\alpha(\hat{\zeta}^\dagger)^\alpha\right\rangle$ can be simplified using the generalized Wick's theorem. Along with the commutation relation $[\widetilde{a}_{as}(L,\omega),\widetilde{a}_{as}^\dagger(L,-\omega')]=\frac{L}{2\pi c}\delta(\omega+\omega')$, this term can be expressed as:
\begin{align}
\left\langle
\hat{\zeta}^\alpha(\hat{\zeta}^\dagger)^\alpha
\right\rangle
=&
\alpha!
\left\langle
\hat{\zeta}\hat{\zeta}^\dagger
\right\rangle^\alpha
=
\alpha!
\left[
 \frac{L}{c}
 \int \frac{d\omega}{2\pi}|B|^2
\right]^\alpha.
\end{align}
Similarly, the operator product term $\left\langle(\hat{\eta}^\dagger)^{l+n-\alpha}\hat{\eta}^{l+n-\alpha}\right\rangle$ can be simplified as follows:
\begin{align}
&
\left\langle
 (\hat{\eta}^\dagger)^{l+n-\alpha}\hat{\eta}^{l+n-\alpha}
\right\rangle
=(l+n-\alpha)!
  \left[
   \int\frac{d\omega}{2\pi}
   \langle \widetilde{F}_s^\dagger(-\omega)\widetilde{F}_s(\omega)\rangle
  \right]^{l+n-\alpha}.
\end{align}
Thus, Eq. (S32) yields
\begin{align}
\rho^s_{nn}
&=
 \sum_{l=0}^{\infty}\frac{(-1)^l}{l!n!}
 \sum_{\alpha=0}^{l+n}
 (C^{l+n}_\alpha)^2
  \alpha!
 \left[
  \frac{L}{c}
  \int \frac{d\omega}{2\pi}|B|^2
 \right]^\alpha
 (l+n-\alpha)!
  \left[
   \int \frac{d\omega}{2\pi}
   \langle \widetilde{F}_s^\dagger(-\omega)\widetilde{F}_s(\omega)\rangle
  \right]^{l+n-\alpha}
%%%%%%%%%%%%%%%%%%%%%%%%%%%%%%%
\nonumber\\&=
 \sum_{l=0}^{\infty}\frac{(-1)^l(l+n)!}{l!n!}
 \sum_{\alpha=0}^{l+n}
 C^{l+n}_\alpha
 \left[
  \frac{L}{c}
  \int \frac{d\omega}{2\pi}|B|^2
 \right]^\alpha
  \left[
   \int \frac{d\omega}{2\pi}
   \langle \widetilde{F}_s^\dagger(-\omega)\widetilde{F}_s(\omega)\rangle
  \right]^{l+n-\alpha}
%%%%%%%%%%%%%%%%%%%%%%%%%%%%%%%
\nonumber\\&=
 \sum_{l=0}^{\infty}\frac{(-1)^l(l+n)!}{l!n!}
 \left[
 \frac{L}{c}
 \int d\omega
  \widetilde{R}_s(\omega)
 \right]^{l+n}
%%%%%%%%%%%%%%%%%%%%%%%%%%%%%%%
=
\frac{\left(\frac{L}{c}R_s\right)^n}
{\left(1+\frac{L}{c}R_s\right)^{n+1}},
\end{align}
where the term $\frac{L}{c}R_s=\langle\hat{a}^\dagger_s(L,t)\hat{a}_s(L,t)\rangle$ represents the average photon number of the Stokes photons $\bar{n}_s$. Hence, the density matrix of the Stokes field becomes
\begin{align}
\rho^s=\sum_{n=0}^\infty
\frac{\bar{n}_s^n}{\left( 1+\bar{n}_s \right)^{n+1}}
|n\rangle\langle n|,
\end{align}
which shows the characteristic of thermal field distribution. Similarly, we can consider the anti-Stokes field operator in the time domain to have the following form:
\begin{align}
\hat{a}^\dagger_{as}(0,t)=&\int d\omega e^{-i\omega t}
\left[
C\widetilde{a}_s(0,\omega)
+D\widetilde{a}_{as}^{\dagger}(L,-\omega)
+\widetilde{F}_{as}(\omega)
\right].
\end{align}
This identical structure enables us to deduce its density matrix, denoted as $\rho^{as}$, utilizing the previously mentioned method. Consequently, we acquire the density matrix for the anti-Stokes field as outlined below:
\begin{align}
\rho^{as}=\sum_{n=0}^{\infty}\frac{\bar{n}_{as}^n}{\left(1+\bar{n}_{as}\right)^{n+1}}
|n\rangle\langle n|,
\end{align}
also exhibiting the characteristics of a thermal distribution.

%%%%%%%%%%%%%%%%%%%%%%%%%%%%%%%%%%%%%%%%%%%%%%%%%%%%%%%%%%%%%%%%%%%%%%%%%%%%%%%%%%%%%%%%%%%%%%%%%%%%%
%%%%%%%%%%%%%%%%%%%%%%%%%%%%%%%%%%%%%%%%%%%%%%%%%%%%%%%%%%%%%%%%%%%%%%%%%%%%%%%%%%%%%%%%%%%%%%%%%%%%%

\section{Experimental Details} 

\subsection{Coincidence Count Rate}

The coincidence count rate characterizes the detection rate in the anti-Stokes channel for a specific number of received Stokes photons. This rate is closely related to the normalized Glauber second-order cross-correlation function, defined as $R_{\rm C}(\tau)=R_sR_{as}g_{s\text{-}as}^{(2)}(\tau)\Delta T$, where $\Delta T$ represents the time bin for detecting the Stokes photons. To ensure a fixed value of unity for the post-selected Stokes photon, we chose a time bin of $\Delta T=1/R_s$. This choice leads us to the following expressions:
\begin{align}
R_{\rm C}(\tau)=&
R_{as}+
\frac{1}{R_s}
\left|
\int \frac{d\omega}{2\pi}
e^{-i\omega\tau}
\left(
B^*D
+\sum_{jk,j'k'}\int_0^Ldz\,
P_{jk}^*\mathcal{D}_{jk^\dagger,j'k'}Q_{j'k'}\right)
\right|^2.
\end{align}
Our experimental approach involves accumulating $2^{16}$ counts in the Stokes channel for each data point. Figures 2(a)--2(d), 3(a), 3(b), 4(a), and 4(b) in the main text present results accumulated over four data points, corresponding to $2^{18}$ receptions in the Stokes channel. The error bars represent the standard deviation across these four data points. By selecting a total of $2^{18}$ receptions in the Stokes channel, each with a purity $P_s$, we can effectively convert the coincidence count rate $R_{\rm C}$ into coincidence counts $N_{\rm C}$ using the following relations:
\begin{align}
N_{\rm C}=&
2^{18}P_s\times R_{\rm C}\eta_{as}\Delta\tau
+
2^{18}P_s\times R_{\rm noise}^{as}\Delta\tau
+
2^{18}(1-P_{s})\times(R_{\rm noise}^{as}+R_{as}\eta_{as})\Delta\tau.
\end{align}
These relations enable us to quantitatively determine the number of coincidence counts based on the coincidence count rate from the reception characteristics $P_s$ in the Stokes channel, the detection efficiency $\eta_{as}$ in the anti-Stokes channel, and the relevant noise rate $R_{\rm noise}^{as}$. The first term in Eq. (S41) represents the detection of biphotons, corresponding to the registration of both Stokes and anti-Stokes photons. This term includes the contribution of the background from the biphoton generation rate. The second and last terms are environmental background, which are respectively caused by the pure and impure detections in the Stokes channel. If Stokes photons are detected, the impure registers in anti-Stokes channel results in an undesired background; this is represented by the second term. By contrast, an uncorrelated background is necessarily the case if the Stokes channel receives noised photons; this is indicated by the last term. Finally, the experimental biphoton coincidence count rate can be expressed as $N_{\rm C}/(2^{18}P_{s}\eta_{as}\Delta\tau)\equiv R_{\rm C}+R_{\rm env}$, where $R_{\rm env}$ represents the count rate of the environmental background.

%%%%%%%%%%%%%%%%%%%%%%%%%%%%%%%%%%%%%%%%%%%%%%%%%%%%%%%%%%%%%%%%%%%%%%%%%%%%%%%%%%%%%%%%%%%%%%%%%%%%%
%%%%%%%%%%%%%%%%%%%%%%%%%%%%%%%%%%%%%%%%%%%%%%%%%%%%%%%%%%%%%%%%%%%%%%%%%%%%%%%%%%%%%%%%%%%%%%%%%%%%%

\subsection{Pairing Ratio}

Pairing ratio is an important quantity that should be considered. Without introducing this concept, it could potentially lead to misinterpretations of the experimental results. In the main text, we discussed two different forms of pairing ratio definitions. It's important to note that these two definitions are equivalent in reality. This core concept arises from the fact that temporal correlation is not inherently present in all generated biphotons; rather, it needs to be acquired by increasing the ensemble density. In the method using $R_{\rm C}$, we calculate the pairing ratio by integrating $R_{\rm C}$ over delay time $\tau$. When the received Stokes photons are normalized to one, only the proportion of $r_p$ exhibits temporal correlation. Consequently, the proportion of anti-Stokes photons correlated in time with this Stokes photons will also be $r_p$, equivalent to the area of $R_{\rm C}$ with background subtracted. However, it’s crucial to establish the upper limit of this integral before deriving the pairing ratio from our experimental results. To address this concern, we pinpoint a specific time, denoted as $t_0$, based on the theoretical $R_{\rm C}$. The selection of $t_0$ ensures that the cross-correlation function $g^{(2)}_{s\text{-}as}(t_0)$ reaches the background value of 1, and $R_{\rm C}(t_0)$ at this time equals the value of $R_{\rm B}$. For instance, we utilized the same parameters as depicted in Fig. 3(b) of the main text, setting $t_0$ at 180 ns, as demonstrated in the subsequent figure presenting the theoretically predicted $R_{\rm C}$. While the choice of $t_0$ was informed by theoretical curves, our preceding experimental results, including those illustrated in Figs. 2, 3, and 4 of the main text, consistently exhibit a robust agreement between our theoretical model and experimental data. Hence, we deem the selection of $t_0$ to be well-founded.

%%%%%%%%
\FigSTwo
%%%%%%%%

The same concept is evident in Eqs. (1) and (2) presented in the main text. Although the biphotons based on SFWM inherently possess frequency anti-correlation, not all generated biphotons exhibit temporal correlation. This characteristic is elucidated in the Stokes photon generation rate [Eq. (1)] and the anti-Stokes photon generation rate [Eq. (2)]. The generated biphotons are categorized into temporally correlated photons [the first terms of Eqs. (1) and (2)] and temporally uncorrelated photons [the second terms of Eqs. (1) and (2)]. The pairing ratio is defined as the ratio of temporally correlated photons to the total generated photons. In our system, characterized by a small ground-state dephasing rate, the results obtained from the Stokes and anti-Stokes channels are nearly identical.

In contrast to the pairing ratio, the heralding efficiency addresses the practical application of collection efficiency~\cite{heralding1S}. It focuses on measuring anti-Stokes photons and determining the proportion heralded by Stokes photons, essentially representing the collection efficiency $\eta_s$ of the Stokes channel~\cite{heralding2S}. For instance, assuming a biphoton generation rate of $R_{\rm B}$ and a pairing ratio of one ($r_p=1$),  the detection rates on both sides, $R_1=R_{\rm B}\eta_s$ and $R_2=R_{\rm B}\eta_{as}$, can be obtained by separately measuring each channel. Through coincidence detection experiments, the temporally correlated biphoton detection rate $R_3=R_{\rm B}\eta_s\eta_{as}$ can be obtained. These temporally correlated photons occupy a fraction of the detected photons in the anti-Stokes channel with a proportion of $\eta_s$. Specifically, when detecting the anti-Stokes channel, only the rate of $R_3$ is temporally correlated, representing a single-photon Fock state, while the remaining rate of $R_2-R_3$ corresponds to a thermal state.

%%%%%%%%%%%%%%%%%%%%%%%%%%%%%%%%%%%%%%%%%%%%%%%%%%%%%%%%%%%%%%%%%%%%%%%%%%%%%%%%%%%%%%%%%%%%%%%%%%%%%
%%%%%%%%%%%%%%%%%%%%%%%%%%%%%%%%%%%%%%%%%%%%%%%%%%%%%%%%%%%%%%%%%%%%%%%%%%%%%%%%%%%%%%%%%%%%%%%%%%%%%

\subsection{Collection Efficiency}

In the Stokes channel, a pinhole with a 60\% transmission rate is positioned in front of our initial etalon. The first etalon itself has a transmission rate of 60\%, while the second etalon has a transmission rate of 75\%. An optical isolator with an 85\% transmission rate is situated between the two etalons. Subsequent to the second etalon, light is gathered using a multi-mode fiber (MMF) boasting an efficiency of 80\%. The MMF is connected to a single-photon counting module (SPCM) with a quantum efficiency of 65\%. The coupling efficiency between the MMF and SPCM is 92\%. Factoring in the transmission rates of other optical components, the overall collection efficiency is approximately 9\%.

The configuration of the anti-Stokes channel mirrors that of the Stokes channel, featuring a 60\% transmission rate pinhole, two etalons, and an optical isolator with an 85\% transmission rate. The final step involves capturing light using an MMF and coupling it into a SPCM with a quantum efficiency of 65\% (coupling efficiency between MMF and SPCM is 92\%). Noteworthy differences lie in the efficiencies of the etalons and MMF. These variations stem from the optical path design, resulting in differences in the guiding beam size compared to the Stokes case. Both the first and second etalons maintain a transmission rate of 60\%, and the MMF efficiency is 70\%. Considering the transmission rates of other optical components, the overall collection efficiency is approximately 6\%.

For a collection time $T_1$, the total number of received Stokes photons during $T_1$ is given by $R_s\eta_sT_1$, where $\eta_s$ denotes the collection efficiency of the Stokes photons. Due to the anti-Stokes collection efficiency $\eta_{as}$ and the pairing ratio $r_p$, only $R_s\eta_sT_1\times\eta_{as}r_p$ Stokes photons are successfully paired with corresponding anti-Stokes photons. On the other hand, when Stokes and anti-Stokes photons from different pairs are detected, their mutual delay times become stochastic, leading to the appearance of a uniform background signal. Within the collection time $T_1$, for each individual detected Stokes photon (total of $R_s\eta_sT_1$ photons), the corresponding anti-Stokes photon detection rate is $R_{as}\eta_{as}$. Consequently, the average background count within each time spacing $\Delta\tau$ can be expressed as $R_sR_{as}\eta_s\eta_{as}T_1\Delta\tau$. We refer this $\Delta\tau$ to the time bin of the detected anti-Stokes photons.

To ensure sufficient counts at low generation rates, we adopted an alternative approach in our measurements, avoiding the fixed collection time $T_1$. Instead, we maintained a constant count of $2^{18}$ in the Stokes channel to ensure an ample number of detections. As a result, the correlated coincidence counts and uncorrelated background counts were adjusted to $2^{18}P_s\eta_{as}r_p$ and $2^{18}P_sR_{as}\eta_{as}\Delta\tau$, respectively. Here, $P_s$ represents the purity of Stokes detection and is calculated as $P_s=\frac{R_s\eta_s}{R_s\eta_s+R_{\rm noise}^s}$, where $R_{\rm noise}^s$ denotes the count rate of the Stokes channel originating from laser leakage, environmental photons, and dark counts from the SPCM. This parameter was determined by conducting single-channel detection on the Stokes channel over a time interval $T_2$, resulting in a count of $R_s\eta_sT_2+R_{\rm noise}^sT_2$. Furthermore, by theoretically calculating the Stokes generation rate of this process, we obtained the Stokes collection efficiency $\eta_s$. Similarly, the collection efficiency $\eta_{as}$ for the anti-Stokes channel was determined using the same methodology, yielding a count of $R_{as}\eta_{as}T_2+R_{\rm noise}^{as}T_2$ within the time interval $T_2$, where $R_{\rm noise}^{as}$ represents the noise count rate of the anti-Stokes channel. By incorporating these experimentally derived collection efficiencies $\eta_s$ and $\eta_{as}$ into the correlated coincidence counts and uncorrelated background counts, respectively, we were able to obtain the experimental pairing ratio and biphoton generation rate.

The collection efficiencies obtained from our experiments using guiding beams differ from those acquired by measuring the Stokes and anti-Stokes channels in conjunction with the theoretical biphoton generation rate. Specifically, the collection efficiencies obtained from the guiding beams for the Stokes and anti-Stokes fields were 9$\%$ and 6$\%$, respectively, while the corresponding values measured from the biphoton experiments were approximately 2$\%$ and 1$\%$, respectively. For convenience, we refer to the method of determining collection efficiency using the guiding beam as the "guiding method." On the other hand, the approach that relies on the measured photon counts from biphoton experiments and the theoretical generation rate is termed the "experimental method." The disparity between these two approaches may arise from differences in the propagation of the guiding beam and the biphoton field, potentially leading to variations in the coupling efficiency with etalons and optical fibers. In this research, we consider the experimental method to be more suitable. This choice is reinforced by the alignment between our experimental results and the theoretical predictions for the signal-to-background ratio and biphoton wavepacket characteristics. Conversely, using the collection efficiencies of 9$\%$ and 6$\%$ obtained from the guiding method would lead to deviations from the theoretically predicted generation rates. This, in turn, would result in differences between the theoretical signal-to-background ratios and biphoton wavepacket characteristics and their experimental counterparts.

%%%%%%%%%%%%%%%%%%%%%%%%%%%%%%%%%%%%%%%%%%%%%%%%%%%%%%%%%%%%%%%%%%%%%%%%%%%%%%%%%%%%%%%%%%%%%%%%%%%%%
%%%%%%%%%%%%%%%%%%%%%%%%%%%%%%%%%%%%%%%%%%%%%%%%%%%%%%%%%%%%%%%%%%%%%%%%%%%%%%%%%%%%%%%%%%%%%%%%%%%%%

\subsection{Biphoton Temporal Profile}

We introduce an alternative method to characterize the generated biphoton by directly calculating the biphoton wave function~\cite{DuS}. The SFWM is a process of third-order susceptibility $\chi^{(3)}$: 
$H_I=\epsilon_0A
\int_{-L/2}^{L/2}dz\chi^{(3)}
E_d^{(+)}E_c^{(+)}\hat{E}_s^{(-)}\hat{E}_{as}^{(-)}+{\rm H.c.}$, where $E_d^{(+)}=E_de^{i(k_dz-\omega_dt)}$ and $E_c^{(+)}=E_ce^{-i(k_cz+\omega_ct)}$ are positive-frequency driving and coupling electric fields, respectively. The generated Stokes field is treated using the method of quantized field with the form $\hat{E}_s^{(-)}=\sqrt{\frac{\hbar\bar{\omega}_s}{2\epsilon_0V}}
e^{-i(k_sz-\bar{\omega}_st)}
\int d\omega_1\widetilde{a}_s^\dagger(-\omega_1)
e^{-i\left[\frac{k_s\chi_s^*(-\omega_1)}{2}z+\omega_1t\right]}$, whereas the anti-Stokes field is described with the form $\hat{E}_{as}^{(-)}=\sqrt{\frac{\hbar\bar{\omega}_{as}}{2\epsilon_0V}}
e^{i(k_{as}z+\bar{\omega}_{as}t)}
\int d\omega_2\widetilde{a}_{as}^\dagger(-\omega_2)
e^{i\left[\frac{k_{as}\chi_{as}^*(-\omega_2)}{2}z-\omega_2t\right]}$. In our calculations, we have supposed $k(\omega)=k_0\sqrt{1+\chi(\omega)}\approx k_0+\frac{\chi(\omega)}{2}$. Upon substituting these electric fields into the interaction Hamiltonian, it can be obtained that
\begin{align}
H_I=&
\frac{\hbar\sqrt{\bar{\omega}_s\bar{\omega}_{as}}}{2L}E_dE_c
\int d\omega_1\int d\omega_2
\int_{-L/2}^{L/2}dz\,e^{i\Delta kz}
e^{-\frac{i}{2}\left[k_s\chi_s^*(-\omega_1)-k_{as}\chi_{as}^*(-\omega_2)\right]z}
e^{-i(\omega_1+\omega_2)t}
\chi^{(3)}(\omega_1,\omega_2)
\widetilde{a}_s^{\dagger}(-\omega_1)\widetilde{a}_{as}^{\dagger}(-\omega_2)
\nonumber\\&+{\rm H.c.}
\nonumber\\
=&%------------------------------------
\frac{\hbar\sqrt{\bar{\omega}_s\bar{\omega}_{as}}}{2L}E_dE_c
\int d\omega_1\int d\omega_2
\int_{-L/2}^{L/2}dz\,e^{i\kappa(\omega_1,\omega_2)z}
e^{-i(\omega_1+\omega_2)t}
\chi^{(3)}(\omega_1,\omega_2)
\widetilde{a}_s^{\dagger}(-\omega_1)\widetilde{a}_{as}^{\dagger}(-\omega_2)+{\rm H.c.}
\nonumber\\
=&%-----------------------------------------
\frac{\hbar\sqrt{\bar{\omega}_s\bar{\omega}_{as}}}{2}E_dE_c
\int d\omega_1\int d\omega_2\,
{\rm sinc}\left[\frac{\kappa(\omega_1,\omega_2)L}{2}\right]
e^{-i(\omega_1+\omega_2)t}
\chi^{(3)}(\omega_1,\omega_2)
\widetilde{a}_s^{\dagger}(-\omega_1)\widetilde{a}_{as}^{\dagger}(-\omega_2)+{\rm H.c.},
\end{align}
where we define $\kappa(\omega_1,\omega_2)=\Delta k-\frac{k_s\chi_s^*(-\omega_1)}{2}+\frac{k_{as}\chi_{as}^*(-\omega_2)}{2}$. The biphoton wave function $|\psi\rangle$ is the solution of the Schrödinger equation: $\frac{\partial}{\partial t}|\psi\rangle=\frac{1}{i\hbar}H_I|\psi\rangle$, which yields the results of
\begin{align}
|\psi(t)\rangle=&|0\rangle
+\frac{1}{i\hbar}\int_{-\infty}^tdt'H_I(t')|\psi(t')\rangle
\nonumber\\
=&|0\rangle
+\frac{1}{i\hbar}\int_{-\infty}^{t}dt'H_I(t')|0\rangle
+\frac{1}{(i\hbar)^2}\int_{-\infty}^{t}dt'
 \int_{-\infty}^{t'}dt''H_I(t')H_I(t'')|\psi(t'')\rangle.
\end{align}
The first term in Eq. (S43) can be safely neglected, as the vacuum field is inherently undetectable. Moreover, we can disregard the integrand $H_I(t')H_I(t'')$ in the last terms, given its minimal impact attributed to $\chi^{(3)}$. Consequently, the expression for the steady-state biphoton wave function $|\psi(t\rightarrow\infty)\rangle$ can be stated as follows:
\begin{align}
|\psi(t\rightarrow\infty)\rangle
=&
-\frac{iE_dE_c\sqrt{\bar{\omega}_s\bar{\omega}_{as}}}{2}
\int_{-\infty}^{\infty}dt'
\int d\omega_1\int d\omega_2\,
{\rm sinc}\left[\frac{\kappa(\omega_1,\omega_2)L}{2}\right]
e^{-i(\omega_1+\omega_2)t}
\chi^{(3)}(\omega_1,\omega_2)
\widetilde{a}_s^{\dagger}(-\omega_1)\widetilde{a}_{as}^{\dagger}(-\omega_2)
|0\rangle
\nonumber\\
=&%-----------------------------------------
-i\pi E_dE_c\sqrt{\bar{\omega}_s\bar{\omega}_{as}}
\int d\omega_1\,
{\rm sinc}\left[\frac{\kappa(\omega_1,-\omega_1)L}{2}\right]
\chi^{(3)}(\omega_1,-\omega_1)
\widetilde{a}_s^{\dagger}(-\omega_1)\widetilde{a}_{as}^{\dagger}(\omega_1)
|0\rangle.
\end{align}
It is worth noting that in the calculation of this biphoton wave function, $\widetilde{a}_s^\dagger$ and $\widetilde{a}_{as}^\dagger$ appear together. This implies that this method can only calculate correlated photons participating in FWM and, as a result, cannot directly provide information about the pairing ratio $r_p$.

By utilizing the steady-state biphoton wave function, we can calculate the Glauber second-order correlation function as $G^{(2)}_{s\text{-}as}(\tau)=\langle\psi(t\rightarrow\infty)|\hat{a}_s^\dagger(t)\hat{a}_{as}^\dagger(t+\tau)\hat{a}_{as}(t+\tau)\hat{a}_s(t)|\psi(t\rightarrow\infty)\rangle$. Since the biphoton wave function is associated with $\widetilde{a}_s^{\dagger}\widetilde{a}_{as}^{\dagger}|0\rangle$, we can insert $|0\rangle\langle0|$ between $\hat{a}_{as}^\dagger(t+\tau)$ and $\hat{a}_{as}(t+\tau)$, where $|0\rangle$ denotes the vacuum state. This yields $G^{(2)}_{s\text{-}as}(\tau)=|\langle0|\hat{a}_{as}(t+\tau)\hat{a}_s(t)|\psi(t\rightarrow\infty)\rangle|^2$. The term $\langle0|\hat{a}_{as}(t+\tau)\hat{a}_s(t)|\psi(t\rightarrow\infty)\rangle$ can be expressed as follows:
\begin{align}
&\langle0|\hat{a}_{as}(t+\tau)\hat{a}_s(t)|\psi(t\rightarrow\infty)\rangle
\nonumber\\
&=
-i\pi E_dE_c\sqrt{\bar{\omega}_s\bar{\omega}_{as}}
\int d\omega_1\,
{\rm sinc}\left[\frac{\kappa(\omega_1,-\omega_1)L}{2}\right]
\chi^{(3)}(\omega_1,-\omega_1)
\langle0|
\hat{a}_{as}(t+\tau)\hat{a}_s(t)
\widetilde{a}_s^{\dagger}(-\omega_1)\widetilde{a}_{as}^{\dagger}(\omega_1)
|0\rangle
\nonumber\\
&=%-------------------------------------------
-i\pi E_dE_c\sqrt{\bar{\omega}_s\bar{\omega}_{as}}
\int d\omega_1\,
{\rm sinc}\left[\frac{\kappa(\omega_1,-\omega_1)L}{2}\right]
\chi^{(3)}(\omega_1,-\omega_1)
e^{i(k_s\frac{L}{2}-\bar{\omega}_st)}
e^{i(k_{as}\frac{L}{2}-\bar{\omega}_{as}t-\bar{\omega}_{as}\tau)}
\nonumber\\&\quad\times
\int d\omega_3
e^{i\left[\frac{k_s\chi_s(\omega_3)}{2}\frac{L}{2}-\omega_3t\right]}
\int d\omega_4
e^{i\left[\frac{k_{as}\chi_{as}(\omega_4)}{2}\frac{L}{2}-\omega_4t-\omega_4\tau\right]}
\langle0|
\widetilde{a}_{as}(\omega_4)\widetilde{a}_s(\omega_3)
\widetilde{a}_s^{\dagger}(-\omega_1)\widetilde{a}_{as}^{\dagger}(\omega_1)
|0\rangle
\nonumber\\
&=%-------------------------------------------
-i\pi E_dE_c\sqrt{\bar{\omega}_s\bar{\omega}_{as}}
\frac{L^2}{4\pi^2c^2}
e^{i(k_s\frac{L}{2}-\bar{\omega}_st)}
e^{i(k_{as}\frac{L}{2}-\bar{\omega}_{as}t-\bar{\omega}_{as}\tau)}
\nonumber\\&\quad\times
\int d\omega_1\,
{\rm sinc}\left[\frac{\kappa(\omega_1,-\omega_1)L}{2}\right]
\chi^{(3)}(\omega_1,-\omega_1)
e^{\frac{i}{2}\left[k_s\chi_s(-\omega_1)+k_{as}\chi_{as}(\omega_1)\right]\frac{L}{2}}
e^{-i\omega_1\tau}
\nonumber\\
&=%--------------------------------------
\frac{-iE_dE_c\sqrt{\bar{\omega}_s\bar{\omega}_{as}}}{2}
\frac{L^2}{c^2}
e^{i\phi(t)}
\int\frac{d\omega_1}{2\pi}e^{-i\omega_1\tau}
{\rm sinc}\left[\frac{\kappa(\omega_1,-\omega_1)L}{2}\right]
\chi^{(3)}(\omega_1,-\omega_1)
e^{\frac{i}{2}\left[k_s\chi_s(-\omega_1)+k_{as}\chi_{as}(\omega_1)\right]\frac{L}{2}},
\end{align}
where $\phi(t)=\frac{k_s+k_{as}}{2}L-\bar{\omega}_{as}\tau-(\bar{\omega}_s+\bar{\omega}_{as})t$. By employing the semiclassical model and assuming a large driving detuning $\Delta_d$, we derive the forms of susceptibilities as follows:
\begin{align}
\chi_s(\omega)=&
\frac{n|d_{32}|^2}{\epsilon_0\hbar}
\frac{|\Omega_d|^2}{\Delta_d^2}
\frac{\omega-i\Gamma/2}{|\Omega_c|^2-4(\omega-i\gamma_{21}/2)(\omega-i\Gamma/2)},
\\
\chi_{as}(\omega)=&
\frac{n|d_{41}|^2}{\epsilon_0\hbar}
\frac{4(\omega-\Delta_c+i\gamma_{21}/2)}{|\Omega_c|^2-4(\omega-\Delta_c+i\gamma_{21}/2)(\omega+i\Gamma/2)},
\\
\chi^{(3)}(\omega)=&
\frac{nd_{14}d_{23}d_{31}d_{42}}{\epsilon_0\hbar^3}
\frac{1}{\Delta_d}
\frac{4}{|\Omega_c|^2-4(\omega-\Delta_c+i\gamma_{21}/2)(\omega+i\Gamma/2)}.
\end{align}
On substituting the $\chi^{(3)}$ into $G_{s\text{-}as}^{(2)}(\tau)$, we have
\begin{align}
G_{s\text{-}as}^{(2)}(\tau)
=A_0
\left|
\int\frac{d\omega}{2\pi}e^{-i\omega\tau}
{\rm sinc}\left[\frac{\kappa(\omega,-\omega)L}{2}\right]
A_1(\omega)
e^{\frac{i}{2}\left[k_s\chi_s(-\omega)+k_{as}\chi_{as}(\omega)\right]\frac{L}{2}}
\right|^2,
\end{align}
where $A_0=\frac{L^4n^2|d_{14}|^2|d_{23}|^2\bar{\omega}_s\bar{\omega}_{as}}
{4c^4\epsilon_0^2\hbar^2}=\frac{L^2}{c^2}\frac{\Gamma^2{\rm OD}^2}{16}$ and $A_1(\omega)=\frac{\Omega_d}{\Delta_d}
\frac{\Omega_c}{|\Omega_c|^2-4(\omega-\Delta_c+i\gamma_{21}/2)(\omega+i\Gamma/2)}$. In the case of a large $\Delta_d$, i.e., $|\chi_s|\ll|\chi_{as}|$, we can further simplify $G_{s\text{-}as}^{(2)}(\tau)$ as follows:
\begin{align}
G_{s\text{-}as}^{(2)}(\tau)
=
\frac{L^2}{c^2}
\left|
\frac{\Gamma{\rm OD}}{4}
\int\frac{d\omega}{2\pi}e^{-i\omega\tau}
{\rm sinc}\left[\frac{\Delta kL}{2}+\frac{k_{as}L\chi_{as}^*(-\omega)}{4}\right]
A_1(\omega)
e^{i\frac{k_{as}L\chi_{as}(\omega)}{4}}
\right|^2.
\end{align}
The coincidence count rate can be determined using the expression $R_{\rm C}(\tau,\Delta T)=\frac{c^2}{L^2}G_{s\text{-}as}^{(2)}(\tau)\Delta T$. Here, we use a time bin of $\Delta T=1$ second to calculate the correlated biphoton generation rate by integrating $R_{\rm C}(\tau)$, resulting in $R_{\rm B}r_p$ (i.e., $\int d\tau R_{\rm C}(\tau,1\,{\rm s})=R_{\rm B}r_p$). However, the formula in Eq. (S50) does not account for accidental receptions, which can result from biphotons originating from different pairs or those that remain uncorrelated.

After obtaining Eq. (S50), we proceed to explore the regime of damped Rabi oscillations~\cite{DuS}. This regime tends to dominate in cases of low OD or when the coupling field strength $\Omega_c$ is large. In this scenario, the term of ${\rm sinc}\left[\frac{\Delta kL}{2}+\frac{k_{as}L\chi_{as}^*(-\omega)}{4}\right]e^{i\frac{k_{as}L\chi_{as}(\omega)}{4}}$ in the integrand of Eq. (S50) can be approximated as ${\rm sinc}(\Delta kL/2)$. To further simplify the expression, we assume $\gamma_{21}=0\Gamma$, leading to the following form:
\begin{align}
G_{s\text{-}as}^{(2)}(\tau)
=&
\frac{L^2}{c^2}
\left|
\frac{\Gamma\Omega_d\Omega_c{\rm OD}}{4\Delta_d}
{\rm sinc}\left(\frac{\Delta kL}{2}\right)
\right|^2
\left|
\int\frac{d\omega}{2\pi}e^{-i\omega\tau}
\frac{1}{|\Omega_c|^2-4(\omega-\Delta_c)(\omega+i\Gamma/2)}
\right|^2
\nonumber\\%-------------------------------
=&
\frac{L^2}{c^2}
\left|
\frac{\Gamma\Omega_d\Omega_c{\rm OD}}{16\Delta_d}
{\rm sinc}\left(\frac{\Delta kL}{2}\right)
\right|^2
\left|
\int\frac{d\omega}{2\pi}e^{-i\omega\tau}
\frac{1}{(\omega+i\gamma_e/4-\Omega_e/2)(\omega+i\gamma_e/4+\Omega_e/2)}
\right|^2
\nonumber\\%-------------------------------
=&
\frac{L^2}{c^2}
\left|
\frac{\Gamma\Omega_d\Omega_c{\rm OD}}{8\Omega_e\Delta_d}
{\rm sinc}\left(\frac{\Delta kL}{2}\right)
\right|^2
e^{-\frac{\Gamma}{2}\tau}
\left|
{\rm sin}\left({\frac{\Omega_e}{2}\tau}\right)
\right|^2.
\end{align}
Here, we introduce the parameters $\gamma_e=\Gamma+i\Delta_c$ and $\Omega_e=\sqrt{|\Omega_c|^2+\Delta_c^2-\Gamma^2/4+i\Gamma\Delta_c}$. Subsequently, we further expand $\Omega_e$ as $\alpha+i\beta$, where both $\alpha$ and $\beta$ are real. As a result, the term $\left|{\rm sin}(\Omega_e\tau/2)\right|^2$ can be expressed as $[{\rm sin}^2(\alpha\tau/2)+{\rm sinh}^2(\beta\tau/2)]$, where ${\rm sinh}x=\frac{e^x-e^{-x}}{2}$. Because the term of ${\rm sin}^2(\alpha\tau/2)$ represents the oscillating behavior, the decay processes can be described by $e^{-\frac{\Gamma}{2}\tau}\frac{(e^{\beta\tau/2}-e^{-\beta\tau/2})^2}{4}$. Furthermore, given that $e^{-\beta\tau/2}$ decays more severely, the overall decay behaviors depend on the term $e^{-(\frac{\Gamma}{2}-\beta)\tau}$, which yields the $1/e$ decay time of $1/(\frac{\Gamma}{2}-\beta)$. Keeping other parameters constant, increasing $\Delta_c$ extends the tail of the biphoton wavepacket. This is because the heightened $\Delta_c$ weakens the coupling between the coupling field and the atoms, making it more challenging for the coupling field to efficiently convert the pre-established spinwave excitation $\widetilde{\sigma}_{21}$ into anti-Stokes photons. Under the condition of $\Delta_c=0$, the $1/e$ decay time and oscillating period are $2/\Gamma$ and $2\pi/\sqrt{|\Omega_c|^2-\Gamma^2/4}$, respectively.

%%%%%%%%
\FigSOne
%%%%%%%%

The group delay time in the steady-state condition can be obtained from $\tau_{\rm EIT}=\frac{L}{{\rm Re}[d\omega/dk]}-\frac{L}{c}=L\frac{d}{d\omega}{\rm Re}[\frac{\omega}{c}\sqrt{1+\chi}]-\frac{L}{c}=\frac{L}{c}{\rm Re}[\sqrt{1+\chi}]+\frac{L}{c}\omega\frac{d}{d\omega}{\rm Re}[\sqrt{1+\chi}]-\frac{L}{c}$. The real part of $\sqrt{1+\chi}$ can be dealt with the following method:
\begin{align}
{\rm Re}[\sqrt{1+\chi}]=&
{\rm Re}\left\lbrace\sqrt{1+{\rm Re}[\chi]+i{\rm Im}[\chi]}\right\rbrace
\nonumber\\=&%-------------------------------
{\rm Re}\left\lbrace
\sqrt{
\sqrt{(1+{\rm Re}[\chi])^2+({\rm Im}[\chi])^2}
\left[
\frac{1+{\rm Re}[\chi]}{\sqrt{(1+{\rm Re}[\chi])^2+({\rm Im}[\chi])^2}}
+i
\frac{{\rm Im}[\chi]}{\sqrt{(1+{\rm Re}[\chi])^2+({\rm Im}[\chi])^2}}
\right]
}
\right\rbrace
\nonumber\\\equiv&%--------------------------
{\rm Re}\left\lbrace
\left[(1+{\rm Re}[\chi])^2+({\rm Im}[\chi])^2\right]^{\frac{1}{4}}
\sqrt{e^{i\theta}}
\right\rbrace
\nonumber\\=&%-------------------------------
\left[(1+{\rm Re}[\chi])^2+({\rm Im}[\chi])^2\right]^{\frac{1}{4}}
{\rm cos}(\frac{\theta}{2})
=
\sqrt{\frac{\sqrt{(1+{\rm Re}[\chi])^2+({\rm Im}[\chi])^2}}{2}+\frac{1+{\rm Re}[\chi]}{2}}.
\end{align}
Here, we aim to analyze the propagation characteristics of the anti-Stokes field. In this context, $\omega$ represents the angular frequency denoted as $\omega_{as}$. Consequently, the group delay time can be expressed as:
\begin{align}
\tau_{\rm EIT}(\omega)=\frac{L}{c}{\rm Re}[\sqrt{1+\chi_{as}(\omega)}]+k_{as}L\frac{d}{d\omega}{\rm Re}[\sqrt{1+\chi_{as}(\omega)}]-\frac{L}{c}.
\end{align}
We consider the specific scenario where $\Delta_c=0\Gamma$ and $\omega=0\Gamma$, allowing us to derive the EIT group delay time $\tau_{\rm EIT}=\frac{\Gamma{\rm OD}}{|\Omega_c|^2}$~\cite{FleischhauerS}. It is important to note that under two-photon resonance conditions ($\omega=\Delta_c$), $\tau_{\rm EIT}$ remains constant regardless of variations in $\Delta_c$. Therefore, relying solely on Eq. (S53), we cannot fully elucidate the influence of $\Delta_c$ on the delay time. 

Figure \ref{fig:S1}(a) shows $\tau_{\rm EIT}$ corresponding to different values of $\Delta_c$ and $\omega$. It is evident from the figure that, regardless of the value of $\Delta_c$, the EIT group delay time is equal to $\frac{\Gamma\,{\rm OD}}{|\Omega_c|^2}$ when the condition $\omega=\Delta_c$ is met, as mentioned above. Note that for larger $\Delta_c$ values near $\omega=\Delta_c$, there is a significant variation in $\tau_{\rm EIT}$, which can lead to significant waveform distortion. Conversely, in the vicinity of $\omega=\Delta_c=0\Gamma$, there is a relatively flat region of $\tau_{\rm EIT}$, allowing waveforms to propagate through the medium without severely distortion. Figure \ref{fig:S1}(b) illustrates the transmission of a Gaussian pulse through the EIT medium. The Gaussian pulse has a $1/e^2$ full width of 400 ns. Although Eq. (S53) states that $\tau_{\rm EIT}$ remains constant at the two-photon resonance condition, the pulse behaviors after passing through the EIT medium are noticeably different for various $\Delta_c$ values. It is observed that the tail of the EIT pulse decreases as $\Delta_c$ increases, especially in the case of $\Delta_c=3\Gamma$. This observation further supports the claim made about the significant variation of $\tau_{\rm EIT}$ in Fig. \ref{fig:S1}(a).

By analyzing the damped Rabi oscillation and the EIT group delay, we can gain insights into the behaviors depicted in Figs. 2(c) and 2(d) of the main text. In Fig. 2(c) of the main text, we introduce $\Delta_c=1\Gamma$, which leads to a reduction in $\tau_{\rm EIT}$, as indicated by the red curve in Fig. \ref{fig:S1}(b). Increasing $\Delta_c$ to $3\Gamma$, as demonstrated in Fig. 2(d) of the main text, further decreases $\tau_{\rm EIT}$ due to the reduced effective OD. However, introducing $\Delta_c=3\Gamma$ simultaneously extends the $1/e$ decay time $1/(\frac{\Gamma}{2}-\beta)$ of the damped Rabi oscillation. As a result, the elongated tail in Fig. 2(d) of the main text is attributed to the regime of damped Rabi oscillation.

%%%%%%%%%%%%%%%%%%%%%%%%%%%%%%%%%%%%%%%%%%%%%%%%%%%%%%%%%%%%%%%%%%%%%%%%%%%%%%%%%%%%%%%%%%%%%%%%%%%%%
%%%%%%%%%%%%%%%%%%%%%%%%%%%%%%%%%%%%%%%%%%%%%%%%%%%%%%%%%%%%%%%%%%%%%%%%%%%%%%%%%%%%%%%%%%%%%%%%%%%%%

\subsection{Experimental Parameters of Figure 4}

The driving Rabi frequency $\Omega_d$ in Fig. 4(c) and 4(d) of the main text is 3$\Gamma$. The other parameters are listed in Table S1.

\begin{table}[h]
\centering
\begin{tabular}{|c|c|c|}
\hline
OD & $\Omega_c$ $(\Gamma)$  & $\Delta_d$ $(\Gamma)$ \\
\hline
20 & 4.0 & 5.0 \\
40 & 5.0 & 8.1 \\
60 & 6.2 & 10.2 \\
80 & 7.2 & 12.0 \\
100 & 8.0 & 13.5 \\
120 & 8.8 & 14.9 \\
\hline
\end{tabular}
\caption{Parameters in Fig. 4(c) and 4(d) of the main text.}
\end{table}

%%%%%%%%%%%%%%%%%%%%%%%%%%%%%%%%%%%%%%%%%%%%%%%%%%%%%%%%%%%%%%%%%%%%%%%%%%%%%%%%%%%%%%%%%%%%%%%%%%%%%
%%%%%%%%%%%%%%%%%%%%%%%%%%%%%%%%%%%%%%%%%%%%%%%%%%%%%%%%%%%%%%%%%%%%%%%%%%%%%%%%%%%%%%%%%%%%%%%%%%%%%

%%%%%%%%%%%%%%%%%%%%%%%%%%%%%%%%%%%%%%%%%%%%%%%%%%%%%%%%%%%%%%%%%%%%%%%%%%%%%%%%%%%%%%%%%%%%%%%%%%%%%
%%%%%%%%%%%%%%%%%%%%%%%%%%%%%%%%%%%%%%%%%%%%%%%%%%%%%%%%%%%%%%%%%%%%%%%%%%%%%%%%%%%%%%%%%%%%%%%%%%%%%

\end{document}